\documentclass[osajnl,twocolumn,showpacs,superscriptaddress,10pt]{revtex4-1} 
\usepackage{amsmath,amssymb,graphicx}
\usepackage{algorithmic,algorithm}
\usepackage{color}
\usepackage{grffile}
\usepackage{subfigure}
\begin{document}

\title{GPU accelerated computation of Polarized Subsurface BRDF for Flat Particulate Layers}

\author{Charly Collin}\email{Charly Collin: charly.collin@bobbyblues.com}
\affiliation{University of Central Florida}

\author{Sumanta Pattanaik}\email{Sumanta Pattanaik: sumant@cs.ucf.edu}
\affiliation{University of Central Florida}

\begin{abstract}
BRDF of most real world materials has two components, the surface BRDF due to the light reflecting at the surface of the material and the subsurface BRDF due to the light entering and going through many scattering events inside the material.
Each of these events modifies light's path, power, polarization state.
Computing polarized subsurface BRDF of a material requires simulating  the light transport inside the material. The transport of polarized light is modeled by the Vector Radiative Transfer Equation (VRTE), an integro-differential equation.
Computing solution to that equation is expensive.
The Discrete Ordinate Method (DOM) is a common approach to solving the VRTE.
Such solvers are very time consuming for complex uses such as BRDF computation, where one must solve VRTE for surface radiance distribution due to light incident from every direction of the hemisphere above the surface.
In this paper, we present a GPU based DOM solution of the VRTE to expedite the subsurface BRDF computation.
As in other DOM based solutions, our solution is based on Fourier expansions of the phase function and the radiance function.
This allows us to independently solve the VRTE for each order of expansion.
We take advantage of those repetitions and of the repetitions in each of the sub-steps of the solution process.
Our solver is implemented to run mainly on graphics hardware using the OpenCL library and runs up to seven times faster than its CPU equivalent, allowing the computation of subsurface BRDF in a matter of minutes.
We compute and present the subsurface BRDF lobes due to powders and paints of a few materials. We also show the rendering of objects with the computed BRDF. The solver is available  for public use through the authors' web site.
\end{abstract}


\maketitle

\section{Introduction}
The propagation of light in scattering and absorbing media is modeled by the Radiative Transfer Equation (RTE) \cite{chandrasekhar}.
One of the most common RTE solution methods is the Discrete Ordinate Method (DOM), which is the method used in publicly available solvers, such as DISORT \cite{disort}.

To take into account polarization nature of the light, the vector radiative transfer equation (VRTE), a variant of the RTE, has been proposed.
To support polarization, the VRTE incorporates Stokes vector representation of polarized light and Mueller matrix representation of polarized BRDF and phase function.
As its scalar equivalent (RTE), the VRTE is an integro differential equation, but rather than working with scalar values, the VRTE is expressed in terms of Stokes vectors and Mueller matrices \cite{hecht2002optics}.
The DOM solution can still be used for the numerical solution of the VRTE \cite{siewert2000jqsrt}, the difference being that the vector problem involves complex arithmetic.
There are a number of references to such DOM based solver in literature.
ARTS \cite{arts} and VLIDORT \cite{vlidort} are two well known public domain DOM based VRTE solvers.

Like many numerical solvers, a DOM VRTE solver is computationally expensive.
As we will notice in later sections, in DOM based solution the order  of computation increases polynomially with the number incident directions. Consequently, for applications involving a large number of incident and outgoing directions, serial implementation of DOM solver can take long computation time (say hours).

In this paper, we describe a GPU based parallel BRDF computation which relies heavily on a DOM solution for VRTE for layered materials.
Our solver running on a off the shelf GPU, carries out the computation in much faster (about seven times) than serial version, and hence allows us to compute a BRDF in a few minutes.

An overview of the BRDF computation and light transport solution is given in section \ref{section_solution}.
Section \ref{section_cost} presents a detailed cost analysis of the VRTE solution.
Section \ref{section_results} describes our GPU solver and presents the computation time as a function of various parameters, follows it with the analysis of the results.

\section{Solution}
\label{section_solution}

Our goal is to compute the subsurface component of the BRDF $\mathbf{F_r}$.
This corresponds to the light entering the material, getting scattered and exiting at the same or at points very close to the point of light entry.
This component satisfies the following equation:
\begin{eqnarray}
\label{eqn_brdf}
\mathbf{F_r}(\mu^\prime, \phi^\prime, \mu, \phi)\mathbf{E}(\mu^\prime, \phi^\prime) = \mathbf{I}(\mu, \phi),
\end{eqnarray}
where $\mu$ is the cosine of the zenith angle and $\phi$ is the azimuthal angle of a direction.
In this equation, $\mathbf{E}(\mu^\prime, \phi^\prime)$  and $\mathbf{I}(\mu, \phi)$ are respectively the Stokes representation of polarized incident irradiance and exiting radiance, and $\mathbf{F_r}$ is the BRDF Mueller matrix.
The components of the Stokes vectors $[I, Q, U, V]$ are: $I$ the total radiance, $Q$  the difference between the linearly polarized components of radiance along the horizontal and vertical axis, $U$ the difference between the linearly polarized components at 45 degrees and 135 degrees and $V$ the difference between the right circularly and left circularly polarized components.

Computing $\mathbf{F_r}$ for a pair of incident and outgoing directions $(\mu^\prime, \phi^\prime, \mu, \phi)$ requires solving the VRTE inside the material to compute the outgoing radiance $\mathbf{I}(\mu, \phi)$ at the surface as a function of incident irradiance from direction $(\mu^\prime, \phi^\prime)$.
In this section we briefly present the VRTE and its DOM solution for layered homogeneous materials and a method for computing the BRDF Mueller matrix from the VRTE solution.

\subsection{VRTE}
Polarized light transport in a plane-parallel participating medium is modeled through the VRTE:
\begin{eqnarray}
\label{eqn_vrte}
\mu \frac{\partial }{\partial \tau}\mathbf{I}(\tau, \mu, \phi) + \mathbf{I}(\tau, \mu, \phi) = \mathbf{J}(\tau, \mu, \phi),
\end{eqnarray}
with $\tau$ the optical depth in the medium.
$\mathbf{I}$ is the polarized radiance and $\mathbf{J}$ is the source term, representing the scattering contribution and is defined as:
{
\small
\begin{eqnarray}
\label{eqn_j}
\mathbf{J}(\tau, \mu, \phi) &=& \frac{\omega(\tau)}{4\pi}\int_{-1}^{1}\int_{0}^{2\pi}\mathbf{P}(\tau, \mu, \phi, \mu^\prime, \phi^\prime)\mathbf{I}(\tau, \mu^\prime, \phi^\prime)d\phi^\prime d\mu^\prime\nonumber\\
&+& \mathbf{Q}(\tau, \mu, \phi),
\end{eqnarray}
}where $\omega$ is the single scattering albedo of the medium, $\mathbf{P}$ the phase matrix, and $\mathbf{Q}$ the inhomogeneous source term which represents the direct contribution from the light source and is defined as:
\begin{eqnarray}
\label{eqn_q}
\mathbf{Q}(\tau, \mu, \phi) = \frac{\omega(\tau)}{4\pi}\mathbf{P}(\tau, \mu, \phi, \mu_0, \phi_0)\mathbf{I_0}\exp^{-\frac{\tau}{\mu_0}},
\end{eqnarray}
with $(\mu_0, \phi_0)$ as the direction towards the light source and $\mathbf{I}_0$ the radiance stokes vector incident from the light source at the top layer.

For layered materials represented by the superposition of homogeneous layers, the phase function and single scattering albedo are unique in each layer and do not depend on $\tau$ anymore.
If we further assume that the layer is particulated and is composed of spherical particles or particles with random azimuthal orientation, then the phase matrix becomes a function of the scattering angle $\theta$ defined for a pair of incident direction $(\mu, \phi)$ and outgoing direction $(\mu^\prime, \phi^\prime)$ as:
\begin{eqnarray}
\label{eqn_scattering_angle}
\theta = \mu \mu^\prime + \sqrt{1 - \mu^2}\sqrt{1 - \mu^{\prime 2}}\cos(\phi - \phi^\prime).
\end{eqnarray}

\subsection{Azimuthal separation}

Because the phase matrix is a function of the scattering angle, its Fourier decomposition can be written as \cite{garcia1996}:
{\small
\begin{eqnarray}
\label{eqn_final_decomposition}
&&\mathbf{P}(\mu, \phi, \mu^\prime, \phi^\prime) = \sum_{m = 0}^{L-1}\sum_{k=1}^2\Phi_k^m(\phi^\prime - \phi)\mathbf{A}^m(\mu, \mu^\prime)\mathbf{D}_k,\\
&&\mathbf{\Phi}_1^m(\phi) = (2 - \delta_{0,m})diag(\cos m\phi, \cos m\phi, \sin m\phi, \sin m\phi),\\
&&\mathbf{\Phi}_2^m(\phi) = (2 - \delta_{0,m})diag(-\sin m\phi, -\sin m\phi, \cos m\phi, \cos m\phi),\nonumber\\
&&\mathbf{D}_1 = diag(1, 1, 0, 0),\\
&&\mathbf{D}_2 = diag(0, 0, 1, 1),\nonumber
\end{eqnarray}
}where the $\mathbf{A}^m$'s are $4\times4$ matrices and are defined as follows \cite{siewert1982}:
\begin{eqnarray}
\label{eqn_phase_function_matrices}
\mathbf{A}^m(\mu, \mu^\prime) = \sum_{l=m}^{L-1}\mathbf{P}_l^m(\mu)\mathbf{B}_l\mathbf{P}_l^m(\mu^\prime).
\end{eqnarray}
Here, $\mathbf{B}_l$ matrices contains the expansion coefficients that define the scattering function and the $\mathbf{P}_l^m$ matrices are the associated Legendre matrices satisfying \cite{siewert1982}:
\begin{eqnarray}
\label{eqn_P_matrices_relation}
&&\mathbf{P}_l^m(-\mu) = (-1)^{l-m}\mathbf{D}\mathbf{P}_l^m(\mu)\mathbf{D},\\
&&\mathbf{D} = diag(1, 1, -1, -1).
\end{eqnarray}
Similarly, the Stokes vectors can be expanded as a Fourier series:
\begin{eqnarray}
\label{eqn_i_expansion}
\mathbf{I}(\tau, \mu, \phi) = \frac{1}{2}\sum_{m = 0}^{L-1}\sum_{k = 1}^2\mathbf{\Phi}_k^m&(\phi - \phi_0)\mathbf{I}_k^m(\tau,\mu).
\end{eqnarray}
Hence, the VRTE can be expanded as a set of equations:
\begin{eqnarray}
\label{eqn_vrte_expanded}
\mu \frac{\partial}{\partial \tau}\mathbf{I}_k^m(\tau, \mu) &=& -\mathbf{I}_k^m(\tau, \mu)\nonumber\\
&+& \frac{\omega}{2}\int_{-1}^1 \mathbf{A}^m(\mu, \mu^\prime)\mathbf{I}_k^m(\tau, \mu^\prime)d\mu^\prime\nonumber\\
&+& \mathbf{Q}_k^m(\tau, \mu),
\end{eqnarray}
where $m \in \{0,L-1\}$, $k \in \{1, 2\}$ and:
{\small
\begin{eqnarray}
\label{eqn_q_expanded}
\mathbf{Q}_k^m(\tau, \mu) &=& \frac{\omega}{4\pi}\mathbf{A}^m(\mu, \mu_0)\mathbf{D}_k\mathbf{I}_0 \exp^{-\frac{\tau}{\mu_0}}.
\end{eqnarray}
}
Using DOM approach, the equation is converted into a discrete set of equations by discretizing $\mu$ using a double Gauss quadrature as follows \cite{stamnes1981}:
\begin{eqnarray}
\label{eqn_vrte_dom}
\mu \frac{\partial}{\partial \tau}\mathbf{I}_k^m(\tau, \mu_i) &=& - \mathbf{I}_k^m(\tau, \mu_i) \nonumber \\
&+& \frac{\omega}{2}\sum_{n = -N}^N \alpha_n
\mathbf{A}^m(\mu_i, \mu_n)\mathbf{I}_k^m(\tau, \mu_n)\nonumber\\
&+& \mathbf{Q}_k^m(\tau, \mu_i),
\end{eqnarray}
where $N$ the quadrature size, $\alpha_n$ are the quadrature weights, and $\mu_i$, $\mu_n$ are the quadrature nodes.
In the rest of this paper, we will assume quadrature nodes are always positive, and will differentiate upward from downward directions by using $\mu_n$ and $-\mu_n$ respectively.

\subsection{Solution}

To compute the exiting radiance field at any optical depth $\tau$, we must solve the equation \ref{eqn_vrte_dom} for all $m$ and $k$ values.
This equation is a first order differential equation and hence its solution can be written as the sum of the homogeneous solution and a particular solution.
In the case of multiple layer materials, a solution has to be computed for each layer. For simplicity, the homogeneous and particular solutions are presented here for a single layer only.
To simplify the equations furthermore, we drop the scripts $k$ and $m$ in the following equations, as the solution presented here is valid for all order of expansion.

\subsubsection{Homogeneous solution}

The homogeneous solution is the solution to the VRTE where the inhomogeneous source term $\mathbf{Q}$ is zero:

{\small
\begin{eqnarray}
\label{eqn_vrte_homogeneous}
\pm\mu_i \frac{\partial}{\partial \tau}\mathbf{I}(\tau, \pm\mu_i) = &-& \mathbf{I}(\tau, \pm\mu_i) \\
&+& \frac{\omega}{2} \sum_{n = 1}^N \alpha_n \mathbf{A}(\pm\mu_i, \mu_n)\mathbf{I}(\tau, \mu_n)\nonumber\\
&+& \frac{\omega}{2} \sum_{n = 1}^N \alpha_n \mathbf{A}(\pm\mu_i,-\mu_n)\mathbf{I}(\tau,-\mu_n).\nonumber
\end{eqnarray}
}Such equations are known to have exponential solutions, so $\mathbf{I}$ can be substituted in \ref{eqn_vrte_homogeneous} as:
\begin{eqnarray}
\label{eqn_I_exponential}
\mathbf{I}(\tau, \pm\mu_i) = \mathbf{\Phi}(\nu, \pm\mu_i)\exp^{-\tau / \nu}.
\end{eqnarray}
This leads to the following set of homogeneous equations:
{\scriptsize
\begin{eqnarray}
\label{eqn_vrte_homogeneous_2}
-\frac{\pm\mu_i}{\nu}\mathbf{\Phi}(\nu, \pm\mu_i) &=& \sum_{n = 1}^N\left(\frac{\omega}{2}\alpha_n\mathbf{A}(\pm\mu_i, \mu_n) - \delta_{i,n}\right)\mathbf{\Phi}(\nu, \mu_n)\\
&+& \sum_{n = 1}^N\left(\frac{\omega}{2}\alpha_n\mathbf{A}(\pm\mu_i, -\mu_n) - \delta_{i,n}\right)\mathbf{\Phi}(\nu, -\mu_n)\nonumber.
\end{eqnarray}
}This set of equations can be written as a matrix operation using the following $4N$ vectors
{\small
\begin{eqnarray*}
\label{eqn_I_vector}
\mathbf{\Phi}^\pm(\nu) = [\mathbf{\Phi}^T(\nu, \pm\mu_1), \mathbf{\Phi}^T(\nu, \pm\mu_2), \cdots, \mathbf{\Phi}^T(\nu, \pm\mu_N)]^T,
\end{eqnarray*}
}
as
\begin{eqnarray}
\label{eqn_eigen_problem}
\mathbf{M}^{-1}\left(\frac{\omega}{2}\mathbf{W}\mathcal{A} - \mathbf{I}_{8N}\right)
 \left[\begin{array}{c}
 \mathbf{\Phi}^+(\nu) \\
 \mathbf{\Phi}^-(\nu)
 \end{array}
 \right]=-\frac{1}{\nu}\left[\begin{array}{c}
 \mathbf{\Phi}^+(\nu) \\
 \mathbf{\Phi}^-(\nu)
 \end{array}
 \right],
\end{eqnarray}
where $\mathbf{I}_{8N}$ is the identity matrix of size $8N$, and
\begin{eqnarray}
\label{eqn_eigen_2}
&\mathbf{M}& = diag(\mu_1\mathbf{I_4}, \cdots, \mu_N\mathbf{I_4}, -\mu_1\mathbf{I_4}, \cdots, -\mu_N\mathbf{I_4}),\\
&\mathbf{W}& = diag(\alpha_1\mathbf{I_4}, \dots, \alpha_N\mathbf{I_4}, \alpha_1\mathbf{I_4}, \dots, \alpha_N\mathbf{I_4}).
\end{eqnarray}
$\mathcal{A}$ in the above equation is a $8N\times8N$ matrix defined as follow:
\begin{eqnarray}
\label{eqn_eigen_A}
\mathcal{A} = \left[
\begin{array}{cc}
\mathcal{A}^1 & \mathcal{A}^2 \\
\mathcal{A}^3 & \mathcal{A}^4 \\
\end{array}
\right],
\end{eqnarray}
where each subblock of the $4N\times4N$ $\mathcal{A}^i$ matrices are:
\begin{eqnarray}
\mathcal{A}^1_{i,j} = \mathbf{A}(\mu_i, \mu_j),
&\mathcal{A}^2_{i,j} = \mathbf{A}(-\mu_i, \mu_j),&
\mathcal{A}^3_{i,j} = \mathbf{A}(\mu_i, -\mu_j),\nonumber \\
&\mathcal{A}^4_{i,j} = \mathbf{A}(-\mu_i,-\mu_j)&.
\end{eqnarray}
The matrix form of the homogeneous equation \ref{eqn_eigen_problem} represents an eigenproblem, the solution to which yields $4N$ eigenvalues $\nu_j$ and eigenvectors $\mathbf{\Phi}^\pm_j$.
The homogeneous solution is then expressed as a linear combination of those solutions:
{\scriptsize
\begin{eqnarray}
\label{eqn_homogeneous_solution}
\mathbf{I}_+^h(\tau) &=& \sum_{j = 1}^{4N}A_j \mathbf{\Phi}^+ (\nu_j) \exp^{\frac{-\tau}{\nu_j}} + B_j \mathbf{\Phi}^- (\nu_j) \exp^{\frac{-(\tau_0 - \tau)}{\nu_j}},\\
\mathbf{I}_-^h(\tau) &=& \mathbf{\Delta}\sum_{j = 1}^{4N}A_j \mathbf{\Phi}^- (\nu_j) \exp^{\frac{-\tau}{\nu_j} }+ B_j \mathbf{\Phi}^+ (\nu_j) \exp^{\frac{-(\tau_0 - \tau)}{\nu_j}},
\end{eqnarray}
}where
{\small
\begin{eqnarray}
\label{eqn_homogeneous_solution2}
&&\mathbf{I}^h_\pm(\tau) = [\mathbf{I}^T(\tau, \pm\mu_1), \mathbf{I}^T(\tau, \pm\mu_2), \cdots, \mathbf{I}^T(\tau, \pm\mu_N)],\\
&&\mathbf{\Delta} = diag(\mathbf{D}, \mathbf{D}, \cdots, \mathbf{D}),
\end{eqnarray}
}and where $\tau_0$ is the full thickness of the medium.
The computation of the unknowns $A_j$'s and $B_j$'s in Equations \ref{eqn_homogeneous_solution} and 25 will be discussed later.

\subsubsection{Particular solution}
A particular solution to VRTE equation \ref{eqn_vrte_dom} can be found by looking for a solution $\mathbf{I}$ similar to the inhomogeneous term (\ref{eqn_q_expanded}).
This latter has the form:
\begin{eqnarray}
\label{eqn_q_particular}
\mathbf{Q}(\tau, \mu) = \mathbf{X}(\mu)\exp^{\frac{-\tau}{\mu_0}}.
\end{eqnarray}
So we look for a solution which can be expressed as:
\begin{eqnarray}
\label{eqn_i_particular}
\mathbf{I}^p(\tau, \mu) = \mathbf{Z}(\mu)\exp^{\frac{-\tau}{\mu_0}}.
\end{eqnarray}
Substituting (\ref{eqn_i_particular}) in (\ref{eqn_vrte_dom}) yields the following set of particular equations:
\begin{eqnarray}
\label{eqn_particular}
-\frac{\pm\mu_i}{\mu_0}\mathbf{Z}(\pm\mu_i) &=& -\mathbf{Z}(\pm\mu_i)\nonumber \\
& + & \frac{\omega}{2}\sum_{n = 1}^N\alpha_n\mathbf{A}(\pm\mu_i, \mu_n)\mathbf{Z}(\mu_n)\nonumber \\
&+& \frac{\omega}{2}\sum_{n = 1}^N\alpha_n\mathbf{A}(\pm\mu_i, -\mu_n)\mathbf{Z}(-\mu_n)\nonumber \\
& + &\mathbf{X}(\pm\mu_i).
\end{eqnarray}
Introducing the $\mathbf{Z}^\pm$ and $\mathbf{X}^\pm$ vectors:
\begin{eqnarray}
\label{eqn_z}
\mathbf{Z}^\pm = [\mathbf{Z}^T(\pm\mu_1), \mathbf{Z}^T(\pm\mu_2), \dots, \mathbf{Z}^T(\pm\mu_N)]^T,\\
\mathbf{X}^\pm = [\mathbf{X}^T(\pm\mu_1), \mathbf{X}^T(\pm\mu_2), \dots, \mathbf{X}^T(\pm\mu_N)]^T,
\end{eqnarray}
This set of equation can be written as:
\begin{eqnarray}
\label{eqn_particular_2}
\left(
-\frac{1}{\mu_0}\mathbf{M} + \mathbf{I}_{8N} - \frac{\omega}{2}\mathbf{W}\mathcal{A}
\right)
\left[\begin{array}{c}
 \mathbf{Z}^+ \\
 \mathbf{Z}^-
 \end{array}
 \right] =
 \left[\begin{array}{c}
 \mathbf{X}^+\\
 \mathbf{X}^-
 \end{array}
 \right].
\end{eqnarray}

Once solved, equation \ref{eqn_particular_2} yields the $\mathbf{Z}^\pm$ vectors and the particular solution.

\subsubsection{Boundary conditions}

Having both homogeneous and particular solutions, the radiance field can be computed at any depth $\tau$ in the layer as:
\begin{eqnarray}
\label{eqn_solution}
\mathbf{I}(\tau, \mu_i) = \mathbf{I}^h(\tau, \mu_i) + \mathbf{I}^p(\tau, \mu_i).
\end{eqnarray}
However the homogeneous solution still contains $8N$ unknown $A_j$'s and $B_j$'s.
These unknowns can be computed from $8N$ equations with known $\mathbf{I}$ values.
The boundary condition of the layer gives us those known values.

\begin{itemize}
	\item Light is incident at the top of the material layer from a single direction, which has already been accounted for as the inhomogeneous term. Thus $\mathbf{I}(0,\mu) = 0$ when $\mu < 0.$
	\item The radiance field at the bottom of the medium is due to the reflection of incident radiance field at the bottom boundary and is governed by $\mathbf{R}$, the base material BRDF:
	\begin{eqnarray}
	\label{eqn_solution_base}
	\mathbf{I}(\tau_0, \mu) = \int_0^1\mathbf{R}(\mu, -\mu^\prime)\mu^\prime\mathbf{I}(\tau_0, -\mu^\prime)d\mu^\prime \nonumber\\
+ \mathbf{R}(\mu, -\mu_0)\frac{\mu_0}{2\pi}\mathbf{I}_0\exp^{-\tau_0 / \mu_0}.
	\end{eqnarray}
The second term on the righthand side of the equation is due to the reflection of the attenuated incident radiance.\\
For a simplifying situation where there is no reflection from the base  $\mathbf{I}(\tau_0,\mu) = 0$.
\end{itemize}
Each of those boundary conditions yields a set of $4N$ linear equations.
The unknowns $A$'s and $B$'s can be computed by solving this linear system of $8N$ equations.
In the case of multiple layer materials, an homogeneous solution is computed at each layer.
Each extra layer adds an extra $8N$ unknown combination factors.
To compute those we use the fact that the radiance field is continuous between two layers, so at each layer boundary we have a set of $8N$ linear equations.

\subsubsection{Reconstruction}

Equation \ref{eqn_solution} gives the solution only at the quadrature angles.
For arbitrary angles, the solution to the VRTE needs to be reconstructed.
This is achieved by following the Source Function Integration technique \cite{chandrasekhar}\cite{thomas2002} which yields for any positive $\mu$ value:
\begin{eqnarray}
\label{eqn_reconstruction}
\mathbf{I}(\tau, \mu) &=& \mathbf{I}(\tau_0, \mu)\exp^{-(\tau_0 - \tau) / \mu}\nonumber\\
& + & \int_\tau^{\tau_0}\mathbf{S}(t, \mu)\exp^{-(t - \tau) / \mu}dt,
\end{eqnarray}
\begin{eqnarray}
\label{eqn_reconstruction_bis}
\mathbf{I}(\tau, -\mu) &=& \int_\tau^{0}\mathbf{S}(t, -\mu)\exp^{-(t - \tau) / \mu}dt,
\end{eqnarray}
where:
\begin{eqnarray}
\label{eqn_reconstruction_S}
\mathbf{S}(\tau, \mu) &=& \frac{\omega}{2}\sum_{i = -N}^N \alpha_n \mathbf{A}(\mu, \mu_i)\mathbf{I}(\tau, \mu_i)\nonumber \\
&+& \mathbf{Q}(\tau, \mu).
\end{eqnarray}
The integral terms in equations \ref{eqn_reconstruction}, \ref{eqn_reconstruction_bis} have a close form solution as all the terms are exponential functions of $\tau$.
\subsection{BRDF Computation}
We wish to compute the BRDF Mueller matrix as defined in equation \ref{eqn_brdf}.
For a given pair of directions $(\mu, \mu_0)$, solving equation \ref{eqn_vrte_dom} results in a single Stokes Vector.
To compute the 16 unknowns of the BRDF Mueller matrix, we need solutions to several incident irradiance Stokes vectors $\mathbf{E}_{inc}$ :
\begin{eqnarray}
\label{eqn_fr_2}
\mathbf{F_r}(\mu, \mu^\prime, \phi - \phi^\prime) &=& [\mathbf{I}^1(0, \mu^\prime, \phi^\prime), \dots, \mathbf{I}^n(0, \mu^\prime, \phi^\prime)]\nonumber\\ &\times&[\mathbf{E}_{inc}^1(\mu, \phi), \dots, \mathbf{E}_{inc}^n(\mu, \phi)]^{-1},
\end{eqnarray}
where $n$ is at least 4 when the incident Stokes vectors are linearly independent.

\section{Computation cost}
\label{section_cost}
In this section we analyze the computation cost associated with each step of the VRTE solution, and identify the possible GPU parallelization steps.
\subsection{Homogeneous solution}
The homogeneous solution is the first step of the VRTE solution.
As it is independent of the incident radiance, it needs to be computed only once per order of expansion.
Homogeneous solution involves solving an eigenproblem.
As can be seem from equation \ref{eqn_eigen_problem}, the problem has a size of $8N\times8N$.
Because of the symmetry of the problem, the eigenproblem size is reduced by half by introducing the following vectors and matrices \cite{siewert2000jqsrt}:
\begin{eqnarray}
\label{eqn_cost_homogeneous_X_Y}
\mathcal{X} = \mathbf{M}(\mathbf{\Phi}^+ + \mathbf{\Phi}^-),\nonumber\\
\mathcal{Y} = \mathbf{M}(\mathbf{\Phi}^+ - \mathbf{\Phi}^-),
\end{eqnarray}
{\scriptsize
\begin{eqnarray}
\label{eqn_cost_homogeneous_E_F}
\mathbf{E} = \left(\mathbf{I}_{4N} - \frac{\omega}{2}\sum_{l = m}^L\mathbf{\Pi}_l\mathbf{B}_l\left[\mathbf{I}_{4N} + (-1)^{l-m}\mathbf{D}\right]\mathbf{\Pi}_l^T\mathbf{W}\right)\mathbf{M}^{-1},\nonumber\\
\mathbf{F} = \left(\mathbf{I}_{4N} - \frac{\omega}{2}\sum_{l = m}^L\mathbf{\Pi}_l\mathbf{B}_l\left[\mathbf{I}_{4N} - (-1)^{l-m}\mathbf{D}\right]\mathbf{\Pi}_l^T\mathbf{W}\right)\mathbf{M}^{-1},
\end{eqnarray}
}where:
\begin{eqnarray}
\mathbf{\Pi} = \left[\mathbf{P}_l^m(\mu_1), \mathbf{P}_l^m(\mu_2), \cdots, \mathbf{P}_l^m(\mu_N)\right]^T
\end{eqnarray}
Using the vectors and matrices introduced above gives us the following two eigenproblems:
\begin{eqnarray}
\label{eqn_cost_homogeneous_eigen}
\left(\mathbf{F}\mathbf{E}\right) \mathcal{X} = \lambda\mathcal{X},\\
\left(\mathbf{E}\mathbf{F}\right) \mathcal{Y} = \lambda\mathcal{Y}.
\end{eqnarray}
Solution of either of these problems gives us the required eigenvectors and eigenvalues. For example, solving equation \ref{eqn_cost_homogeneous_eigen} we can retrieve the solution to equation \ref{eqn_eigen_problem} as:
\begin{eqnarray}
\label{eqn_cost_homogeneous_solution}
\nu_j^2 &=& 1 / \lambda_j, \nonumber \\
\mathbf{\Phi}^{\pm}(\nu_j) &=& \frac{1}{2}\mathbf{M}^{-1}(\mathbf{I}_{4N} \pm \nu_j\mathbf{E})\mathcal{X}(\lambda_j).
\end{eqnarray}

While reducing the size of the eigen solution implies an extra eigen solution reconstruction step, it is overall less expensive to do so.

Thus the homogeneous solution is carried out in 3 steps:
\begin{itemize}
\item First the $\mathbf{E}$ and $\mathbf{F}$ matrices are computed. Each matrix is computed one $4\times4$ subblock at a time, each requiring 3 matrix multiplications. This step involves $N\times N\times L$ independent subblock computations that can be computed in parallel.
\item The second step is to solve $L$ eigenproblems, one for each order of expansion. This step also requires that the matrix FE is set up before carrying out the solution. So this step requires $L$ matrix multiplications to compute the $\mathbf{F}\mathbf{E}$ matrices, where dimension of each matrix is $4N \times 4N$. The matrix computation can be done in $L \times 4N \times 4N$ parallel computation $4N$ sum of elements.
\item The final step is to compute the eigenvectors $\mathbf{\Phi}^\pm$ from the half-problem solution. This step involves $L\times4N\times4N$ sums of $4N$ elements, that also can be computed in parallel.
\end{itemize}

\subsection{Particular solution}

The particular solution requires solving $8N$ equations for $8N$ unknowns.
This solution requires solving a linear system of $8N$ equations. Using a strategy similar to that done is homogeneous solution, the problem size is reduced \cite{siewert2000jqsrt} to solving  $4N$ equations for $4N$ unknowns as follows.
This is done by introducing the following vector:
\begin{eqnarray}
\label{eqn_cost_particular_G}
\mathbf{G} = \mathbf{Z}^+ + \mathbf{Z}^-,
\end{eqnarray}
and solving a system of $4N$ linear equations:
\begin{eqnarray}
\label{eqn_cost_particular_equation}
\left(\mathbf{F}\mathbf{E} - \frac{1}{\mu_0^2}\right)\mathbf{G} = \left(\mathbf{F}\mathbf{X}^+ + \frac{1}{\mu_0}\mathbf{X}^-\right)\mathbf{M}^{-1}.
\end{eqnarray}
Once solved, the $\mathbf{Z}^\pm$ vectors as used in (\ref{eqn_z}) can be retrieved as:
\begin{eqnarray}
\label{eqn_cost_particular_Z}
\mathbf{Z}^\pm = \frac{1}{2}\mathbf{M}^{-1}\left[\mathbf{I}_{4N} \pm \frac{1}{\mu_0}\mathbf{E}\right]\mathbf{G}.
\end{eqnarray}

Here are the steps for computing the particular solution:
\begin{itemize}
\item Creation of the problem matrix (equation \ref{eqn_cost_particular_equation}): It involves, for each order of expansion, two $4N\times4N$ matrix-vector multiplications, and thus a total of $2L$ such multiplications. This step can be executed as $L\times4N$ parallel computations, each involving a $4N$ sum and two scalar multiplications.
\item Computing the $\mathbf{G}$ vectors: It requires a total of $L$ $4N \times 4N$ matrix inversions and multiplications.
\item Finally, retrieving the  $\mathbf{Z}^\pm$ vectors: It costs a total of $L$ multiplications of $4N\times4N$ matrices (to compute $\mathbf{E}\mathbf{G}$), as well as $2L$ sums of $4N$ vectors to retrieve $\mathbf{Z}^+$ and $\mathbf{Z}^-$.
\end{itemize}

\subsection{Boundary conditions}
The boundary conditions are used to find the $8N$ unknowns $A$'s and $B$'s for equation \ref{eqn_homogeneous_solution}.
In the simple case of zero reflection from the bottom, the boundary condition leads to the following two sets of equations:
\begin{eqnarray}
\label{eqn_cost_boundary_problem_bb}
&&\mathbf{I}_-^{hT}(0) + \mathbf{I}_-^{pT}(0) = \mathbf{0}_{4N},\\
\label{eqn_cost_boundary_problem_bb_2}
&&\mathbf{I}_+^{hT}(\tau_0) + \mathbf{I}_+^{pT}(\tau_0) = \mathbf{0}_{4N},
\end{eqnarray}
where $\mathbf{0}_{4N}$ is a vertical zero vector of size $4N$.
Introducing the following vector of unknown constants:
\begin{eqnarray}
\label{eqn_cost_boundary_constant_vector}
\mathbf{C}^T = [A_1, A_2, \cdots, A_{4N}, B_1, B_2, \cdots, B_{4N}],
\end{eqnarray}
we can write any homogeneous solution as:
\begin{eqnarray}
\label{eqn_cost_boundary_homogeneous}
\mathbf{I}_h(\tau, \pm\mu_i) = \mathbf{K}^T(\tau, \pm\mu_i)\mathbf{C},
\end{eqnarray}
with:
\begin{eqnarray}
\label{eqn_cost_boundary_k_vector_positive}
\mathbf{K}(\tau, \pm\mu_i) = \left[
                            \begin{array}{c}
                            \mathbf{\Phi}^T(\nu_1, \pm\mu_i)\exp^{\frac{-\tau}{\nu_1}}\\
                            \vdots\\
                            \mathbf{\Phi}^T(\nu_{4N}, \pm\mu_i)\exp^{\frac{-\tau}{\nu_{4N}}}\\
                            \mathbf{\Phi}^T(\nu_1, \mp\mu_i)\exp^{\frac{-(\tau_0-\tau)}{\nu_1}}\\
                            \vdots\\
                            \mathbf{\Phi}^T(\nu_{4N}, \mp\mu_i)\exp^{\frac{-(\tau_0-\tau)}{\nu_{4N}}}
                            \end{array}
                            \right].
\end{eqnarray}
Therefore the boundary problem can be expressed using the following matrix notation:
\begin{eqnarray}
\label{eqn_cost_boundary_bb_final_zero}
\left[
\begin{array}{c}
\mathbf{K}_+^T(0)\\
\mathbf{K}_-^T(\tau_0)
\end{array}
\right]
\mathbf{C}
 =
-\left[
\begin{array}{c}
\mathbf{I}_-^p(0)\\
\mathbf{I}_+^p(\tau_0)
\end{array}
\right],
\end{eqnarray}
where
\begin{eqnarray}
\label{eqn_cost_boundary_bb_k}
&&\mathbf{K}_+(\tau) = [\mathbf{K}(\tau, -\mu_1), \cdots, \mathbf{K}(\tau, -\mu_N)],\\
&&\mathbf{K}_-(\tau) = [\mathbf{K}(\tau,  \mu_1), \cdots, \mathbf{K}(\tau,  \mu_N)].
\end{eqnarray}

Thus in the simple case of no reflection from the bottom boundary, the solution can be split into the following two steps:
\begin{itemize}
\item The creation of the left hand side matrix (equation \ref{eqn_cost_boundary_bb_final_zero}) involves, for each order of expansion, $8N\times8N$ scalar operations.
\item Computing the $\mathbf{C}$ vectors requires a total of $L$ matrix inversions.
\end{itemize}

\subsubsection{Base Reflection}
For solution with nonzero base boundary reflection the righthand side of the equation \ref{eqn_cost_boundary_problem_bb_2} is nonzero and hence we have to take into that computation into account. We rewrite equation \ref{eqn_cost_boundary_problem_bb_2} as
\begin{eqnarray}
&&\mathbf{I}_+^{hT}(\tau_0) + \mathbf{I}_+^{pT}(\tau_0) = \mathbf{L}_{refl}.
\end{eqnarray}
$\mathbf{L}_{refl}$ is derived by discretizing the integral part of the righthand side of equation \ref{eqn_solution_base} for every quadrature angle $\mu_i$ as: 
\begin{eqnarray*}
\mathbf{L}_{refl}(\mu_i) = \sum_{j = 1}^N\alpha_j\mathbf{R}(\mu_i, -\mu_j)\mu_j\mathbf{I}(\tau_0, -\mu_j) + \mathbf{L}_{refl}^s(\mu_i),
\end{eqnarray*}
where $\mathbf{R}$ is the $4\times4$ Mueller BRDF matrix, and $\mathbf{L}_{refl}^s$ represents the reflection of the attenuated source term in equation \ref{eqn_solution_base},
i.e.
\begin{equation}
\label{eqn_cost_boundary_bb_s}
\mathbf{L}_{refl}^s(\mu_i)=\mathbf{R}(\mu_i, -\mu_0)\frac{\mu_0}{2\pi}\mathbf{I}_0\exp^{-\tau_0 / \mu_0}.
\end{equation}
Using equation \ref{eqn_cost_boundary_homogeneous} we rewrite the reflection equation as:
\begin{eqnarray}
\mathbf{L}_{refl}(\mu_i) = \mathbf{L}_{refl}^h(\mu_i)\mathbf{C}+\mathbf{L}_{refl}^{p}(\mu_i)+\mathbf{L}_{refl}^{s}(\mu_i),
\end{eqnarray}
where
{\small
\begin{eqnarray}
\label{eqn_cost_boundary_bb_h}
\mathbf{L}_{refl}^h(\mu_i) &=&
\sum_{j = 1}^N\alpha_j\mathbf{R}(\mu_i, -\mu_j)\mu_j\mathbf{K}^T(\tau_0, -\mu_j),\\
\label{eqn_cost_boundary_bb_p}
\mathbf{L}_{refl}^{p}(\mu_i) &=& \sum_{j = 1}^N\alpha_j\mathbf{R}(\mu_i, -\mu_j)\mu_j\mathbf{I}_p(\tau_0, -\mu_j).
\end{eqnarray}
}Introducing the vectors:
\begin{eqnarray}
\label{eqn_cost_boundary_bb_vectorh}
\mathbf{L}_{refl}^H = \left[
\mathbf{L}_{refl}^{hT}(\mu_1), \cdots, \mathbf{L}_{refl}^{hT}(\mu_N)
\right]^T,\\
\label{eqn_cost_boundary_bb_vectorp}
\mathbf{L}_{refl}^{P} = \left[
\mathbf{L}_{refl}^{pT}(\mu_1), \cdots, \mathbf{L}_{refl}^{pT}(\mu_N)
\right]^T,\\
\label{eqn_cost_boundary_bb_vectors}
\mathbf{L}_{refl}^{S} = \left[
\mathbf{L}_{refl}^{sT}(\mu_1), \cdots, \mathbf{L}_{refl}^{sT}(\mu_N)
\right]^T,
\end{eqnarray}
we now write the boundary problem in the case of non-zero base reflection as:
{\tiny
\begin{eqnarray}
\label{eqn_cost_boundary_bb_final}
\left(
\left[
\begin{array}{c}
\mathbf{K}_+^T(0)\\
\mathbf{K}_-^T(\tau_0)
\end{array}
\right]
-
\left[
\begin{array}{c}
\mathbf{0}_{4N}\\
\mathbf{L}_{refl}^H
\end{array}
\right]
\right)
\mathbf{C}
 =
 \left[
 \begin{array}{c}
 \mathbf{0}_{4N}\\
 \mathbf{L}_{refl}^{P}+\mathbf{L}_{refl}^{S} \end{array}
 \right]
-\left[
\begin{array}{c}
\mathbf{I}_-^p(0)\\
\mathbf{I}_+^p(\tau_0)
\end{array}
\right].
\end{eqnarray}
}

Thus base reflection adds an extra computation step to the boundary problem where the downward radiance field needs to be integrated at the bottom of the layer, and then used to update the boundary condition. The computation of equations \ref{eqn_cost_boundary_bb_h} and \ref{eqn_cost_boundary_bb_p} for every $\mu$ require summation involving $4N$ terms. However, equation \ref{eqn_cost_boundary_bb_h} is evaluated $4N$ times for each order, so a total of $L\times4N$ summations, whereas the equation \ref{eqn_cost_boundary_bb_p} is evaluated only once. So in total this step can be computed as $L\times 4N+1$ parallel summation evaluations. The base reflection also evaluates  equation \ref{eqn_cost_boundary_bb_s}, $4N$ times. However, this evaluation is a simple and hence is not included in this discussion.

In the context of base reflection, the particular case of ideal depolarizing Lambertian base needs a special mention, because it reduces to computation cost to some extent. For such a reflection  only the top-left element of Mueller BRDF matrix $\mathbf{R}$ is non-zero, i.e.
\begin{eqnarray}
\label{eqn_cost_boundary_problem_lb_R}
\mathbf{R}_{00}(\mu_i, -\mu_j) = 2\rho,
\end{eqnarray}
where $\rho$ is the Lambertian albedo.

Furthermore, because of the directional independence of Lambertian reflectors $\mathbf{L}_{refl}^h$, $\mathbf{L}_{refl}^{p}$ and $\mathbf{L}_{refl}^{s}$ in equations \ref{eqn_cost_boundary_bb_h}, \ref{eqn_cost_boundary_bb_p}, and \ref{eqn_cost_boundary_bb_s} are independent of $\mu_i$, hence are computed only once, and are duplicated to set up the corresponding vectors. Thus the computation cost is reduced by a factor of N.

\subsection{Reconstruction}
The reconstruction of the radiance field at any $\tau$ value requires the evaluation of equations \ref{eqn_reconstruction} and \ref{eqn_reconstruction_bis} for any arbitrary $\pm\mu$ value where $\mu\in\{0,1\}$, and for any azimuth angles $\phi$. However, for BRDF computation the radiance field computation is limited only to $\tau=0$ and to $+\mu$ values. However, this computation must be carried out for all incident $\mu_0$ values. Independent of whether it is done for radiance field or for BRDF, the reconstruction step consists of the following four steps.
\begin{itemize}
\item Computing $\mathbf{I}(\tau_0, \mu)$: It is the radiance at the base of the layer (the first item on the right of equation \ref{eqn_reconstruction}). The cost of this step depends on the cost of the bottom boundary condition. In the simplest case where no light is reflected at the bottom of the boundary, that step has no cost as $\mathbf{I}(\tau_0, \mu)$. Otherwise, for every $\mu$ the the cost is a sum of $4N \times 4N$ terms involving homogeneous solution and a sum of $4N$ terms involving particular solution. For Lambertian base, this cost remains the same. However, the term is independent of $\mu$, so needs to be computed only once.
\item Computing $\mathbf{I}$ for equation \ref{eqn_reconstruction_S}:
    We analyze the cost in the following three substeps. Each of these substeps require $N^2$ matrices $\mathbf{A}(\mu, \mu^\prime)$ each of size $4\times 4$, which must be obtained through a sum of up to $L$ elements (see equation \ref{eqn_phase_function_matrices}) each involving two $4\times4$ matrix multiplications.
    \begin{itemize}
    \item Computing ${\bf I}^h$ component of $\mathbf{I}$: It requires $L\times8N$ sums of $2N$ elements, each element of which is obtained through a $4\times 4$ matrix ($\mathbf{A}(\mu, \mu^\prime)$) and $4$ elements vectors ($\mathbf{\Phi}(\nu, \mu)$) multiplication. 

    \item Computing ${\bf I}^p$ component of $\mathbf{I}$: This step requires $L$ sums of $2N$ elements, each element of which is obtained through a $4\times 4$ matrix ($\mathbf{A}(\mu, \mu^\prime)$) and $4$ elements vectors ($\mathbf{Z}(\mu)$) multiplication.

    \item Computing $Q$ term for equation \ref{eqn_reconstruction_S} : It is the last term of equation \ref{eqn_reconstruction_S}. That terms needs to be evaluated at each order of expansion, and for every incident direction in the desired BRDF. For each of these direction, this step can be executed as a single $4\times 4$ matrix ($\mathbf{A}(\mu, \mu_0)$) and $4$ elements vectors ($\mathbf{I}_0$) multiplication. Note that this substep requires $\mathbf{A}(\mu, \mu_0)$ which is different from $\mathbf{A}(\mu, \mu^\prime)$ and hence must be computed independently from the other two substeps.
    \end{itemize}
\end{itemize}

This reconstruction is done for every order of expansion, resulting in all the $\mathbf{I}^m(\tau, \mu)$.
A final reconstruction step is needed to get the radiance at any azimuthal angle $\phi$ using the sum from equation \ref{eqn_i_expansion}.

\section{Implementation, Results and Discussion}
\label{section_results}
We implemented our VRTE solver in C++. For all the parallelizable steps of the solver (that were identified in section 3) 
we created sequential and parallel versions. The parallel version is written in OpenCL (a parallel computing language for multi core and many core devices) \cite{OpenCL}. We chose between the sequential and parallel version through compiler directives. Instead of writing our own  code for Eigen solution (used for Homogeneous solution) and matrix inversion (used for Particular solution and Boundary condition based solution), we used function calls from one of the two well established Linear Algebra C++ libraries: EIGEN \cite{Eigen}, a library that relies on sequential computation, and MAGMA \cite{Magma}, a GPU accelerated library. 
Even though EIGEN is designed for sequential computation, it has an optional vectorization feature that could be enabled to take advantage of the vector instruction set of the CPU.
Taking combinations of sequential and parallel implementation, EIGEN libray (vectorized and nonvectorized) and MAGMA library, we created six different configurations for our solver. These configurations are identified in Table 1. 
In the rest of this section we compare the computation times from these configurations, show the lobes of some of the computed BRDFs, and show some rendered images of a simple scene with objects having the computed BRDFs as their surface reflection properties. The phase functions used in this section (except for the benchmark test) are all computed using a publicly available Mie theory based code by Mischenko \cite{mischenko}.  All the results shown in this section were obtained using a laptop with a Intel Core i7 as CPU and a Nvidia GTX 580 as GPU.

\begin{table}[ht!]
\label{table_configuration}
\centering
\begin{tabular}{|l|c|c|c|}
\hline
  & \begin{tabular}{@{}c@{}c@{}}EIGEN \\{\small without}\\{\small vectorization}\end{tabular} & \begin{tabular}{@{}c@{}c@{}}EIGEN \\{\small with}\\{\small vectorization}\end{tabular}  & \quad MAGMA\quad\ \  \\
\hline
Sequential solution & Conf. 1 & Conf. 2 & Conf. 3\\
\hline
Parallel solution & Conf. 4 &  Con.n 5 & Conf. 6\\
\hline
\end{tabular}
\caption{Various configurations of our VRTE solver.}
\end{table}

\subsection{Validation}
To validate our implementation, we ran the Wauben and Hovenier benchmark test \cite{Hovenier} on all the configurations of our solver. The benchmark uses $30$ quadrature angles and $12$ orders of expansion for the phase function.

The results from all the solver configurations are in perfect agreement with the benchmark results. 
Table 2 
shows the computation time for running the benchmark on each of the solver configurations. The table shows a reasonably good improvement ($\approx 3$ times) between the parallel solver and the sequential solver. Among the configurations, we noticed only minor improvements for using vectorized EIGEN library as compared to using its nonvectorized counterpart. The results from using MAGMA library showed a little discouraging slow down as compared to the computation time from using EIGEN library. We investigate this problem later in the following paragraphs.

\begin{table}[ht!]
\centering
\begin{tabular}{|l|c|c|c|}
\hline
  & \begin{tabular}{@{}c@{}c@{}}EIGEN \\{\small without}\\{\small vectorization}\end{tabular} & \begin{tabular}{@{}c@{}c@{}}EIGEN \\{\small with}\\{\small vectorization}\end{tabular}  & \quad MAGMA\quad\ \  \\
\hline
Sequential solution & $2.79$ & $2.74$& $2.94$\\
\hline
Parallel solution & $0.99$ & $0.97$ & $1.18$ \\
\hline
\end{tabular}
\caption{Time (in seconds) required to compute the benchmark problem using each of the six configurations.}
\label{BenchmarkTimesTable}
\end{table}

\subsection{Radiance field due to a directional incidence}
We ran our solver to compute radiance field due to a directional ($\mu_0=0.6$) monochromatic ($\lambda=400$nm) light incident on  homogeneous particle layers of thickness 100 ($\tau_0=100$) composed of spherical particles (radius $0.6\mu m$) from several materials. We used $40$ quadrature nodes ($N=40$) for the computation. The radiance field was computed for $11$ zenith and $19$ azimuth angles.

\begin{figure}[h]
\centering
\includegraphics[width=0.43\textwidth]{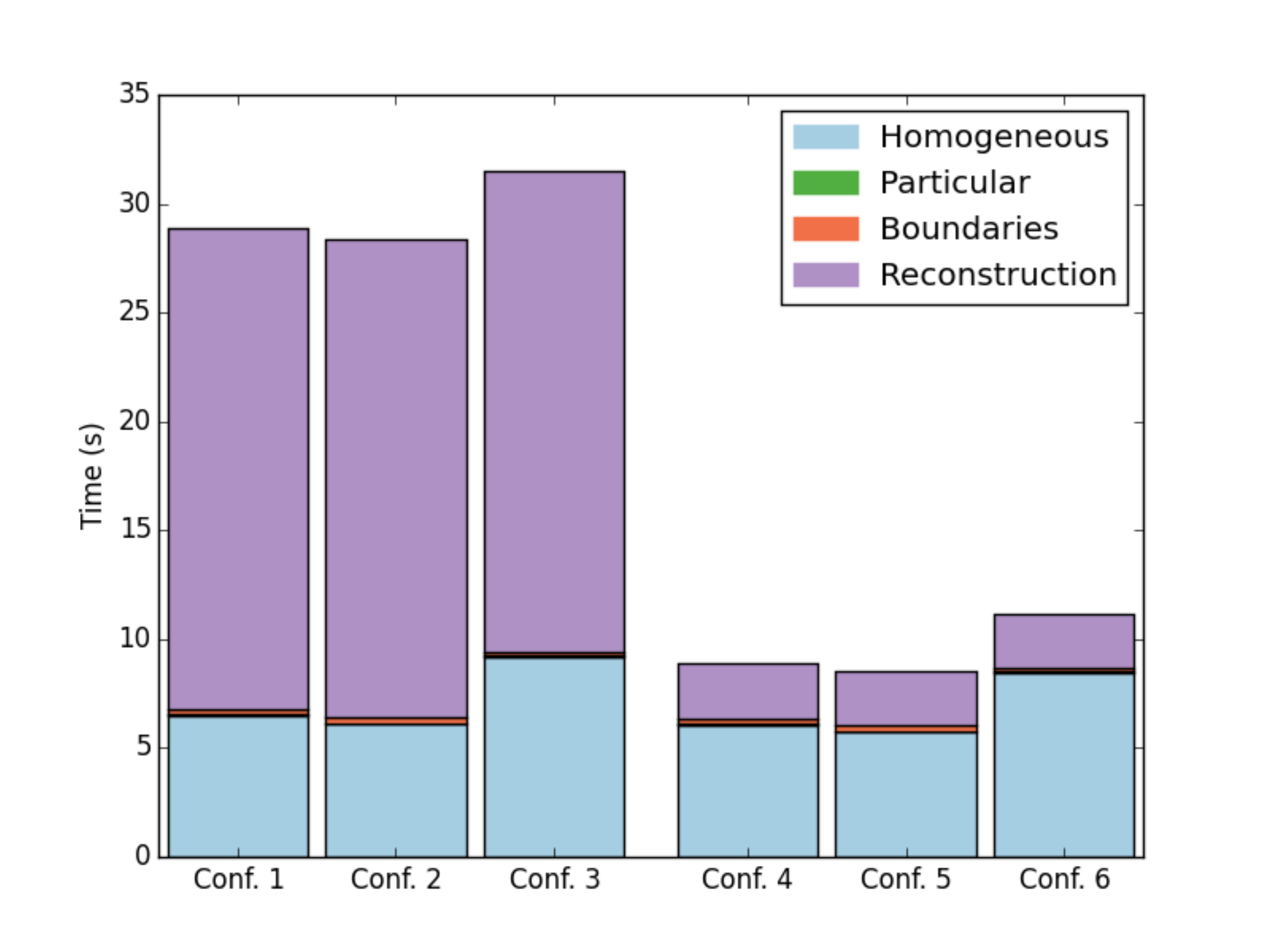}
\caption{Detailed cost of computing a radiance field with the different configurations.}
\label{bar_results_benchmark_times_detailed}
\end{figure}

Table 4 
shows the total computation time
and time spent for each major step of the computation on each of the six configurations for one of the materials (a layered suspension of gold particles). All other materials showed similar trends.
Figure \ref{bar_results_benchmark_times_detailed} shows the same information in a stacked bar chart. We summarize below our observations from the table:
\begin{itemize}
  \item Total computation cost: All the solver configurations that included parallelization ran faster than their sequential counter part. We find a speed-up of more than $3$ between the fastest ones among the sequential and parallel configurations.
  \item Cost of individual computation steps: The computation times for the individual steps do not all show an obvious trend. Independent of the solver configuration, we find that a lion's share of computation time is taken up by the reconstruction step. And this step is also the step that takes the maximum advantage (hence most speedup) from parallelization. If we ignore the configuration using MAGMA (discussed later), the remaining three steps of the computation (homogeneous, particular and boundary solution) do not show any significant speed difference. We justify this finding by pointing to the fact that independent of the configurations, the most expensive matrix computations (eigen solution and matrix inversion) for these steps are done sequentially for each order of expansion using calls to EIGEN library. The particular speed advantage associated with the computation of the matrices in parallel was mostly counter balanced by the time taken to transfer the data to and from the GPU device and in preparing the parameters for EIGEN function calls from the retrieved data. Thus the reconstruction step fully influenced the total speed up in computation time.
  \item Vectorized {\em vs.} nonvectorized EIGEN library: The vectorized EIGEN library consistently showed a slight improvement in computation time over the nonvectorized counterpart. Thus it was unnecessary to continue experimenting with the configurations that used nonvectorized EIGEN library. So we drop them from all the subsequent tables and plots.
 \item EIGEN library {\em vs.} MAGMA library: We observed in our benchmark validation step that the timing results from configurations using MAGMA library has a minor slow down. We notice the same trend here as well. We had expected that the GPU based MAGMA library would improve our computation time for eigen solution (homogeneous solution step) and for matrix inversion (particular solution and boundary handling step). We later found from studies conducted by independent researches \cite{magma_benchmark} that though the GPU based matrix libraries (such as MAGMA) are supposed to accelerate the computation, for complex matrix related problems (such as eigen solution) initial setup times are high and noticeable speedup is only achieved for matrix sizes many time larger than those used in our VRTE solver. 

So we discontinued the configurations using MAGMA library.
 \end{itemize}
 Based on the above observations we continued our experiment using only two remaining configurations: Sequential implementation using vectorized EIGEN library (Configuration 2), and Parallel implementation using vectorized EIGEN library (Configuration 5).

We then went on to study the computation time trend of our configurations as a function of the various parameters of the solver: number of quadrature nodes, orders of expansion, number of outgoing directions for reconstruction. We carried out these experiments to further validate and to verify the scalability of our implementation.  For each of the cases we measured total time and time spent by the individual steps.

\begin{table}[ht!]
\centering
\begin{tabular}{|l|c|c|c|c|c|c|}
\hline
Configuration: & 1 & 2 & 3 & 4 & 5 & 6\\
\hline
Homogeneous: & $6.50$ & $6.11$ & $9.18$ & $6.07$ & $5.75$ &  $8.43$\\
\hline
Particular: & $0.04$ & $0.04$ & $0.07$ & $0.04$ & $0.04$ & $0.07$ \\
\hline
Boundary problem: & $0.24$ & $0.24$ & $0.25$ & $0.24$ & $0.24$ & $0.25$ \\
\hline
Reconstruction: & $22.1$ & $22.0$ & $22.1$ & $2.52$ & $2.51$ & $2.53$ \\
\hline
Total: & $28.9$ & $28.4$ & $31.5$ & $8.88$ & $8.56$ & $11.18$ \\
\hline
\end{tabular}
\caption{Time (in seconds) to solve each step of the computation of the radiance field with the different configurations.}
\label{table_results_single}
\end{table}

\subsubsection{Influence of the quadrature size}

\begin{figure}[h]
\centering
\includegraphics[width=0.43\textwidth]{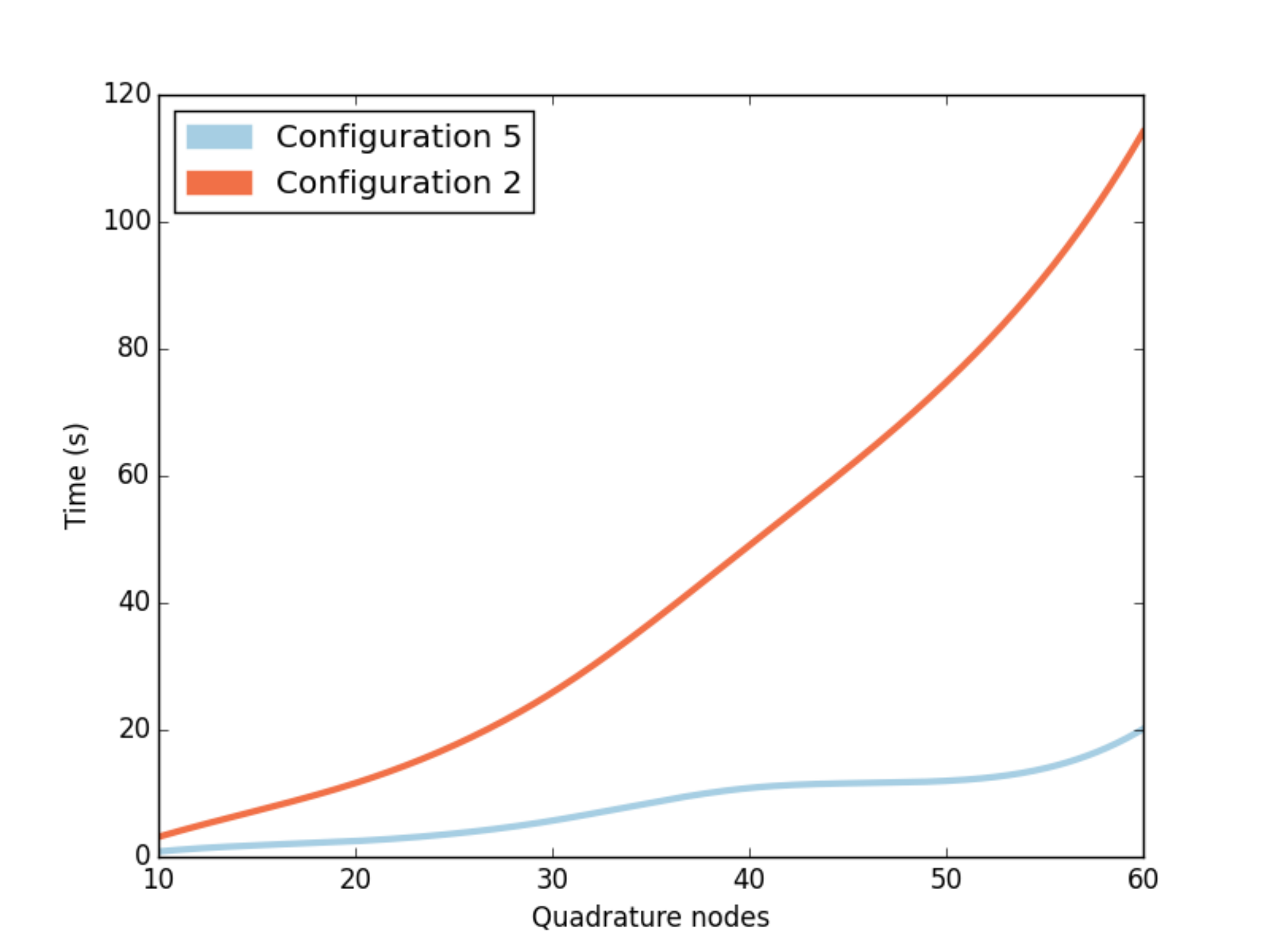}
\caption{Cost of the solution as a function of the quadrature size.}
\label{img_quadrature_1}
\end{figure}

\begin{figure}[h]
\centering
\subfigure[Homogeneous]{\includegraphics[width=0.20\textwidth]{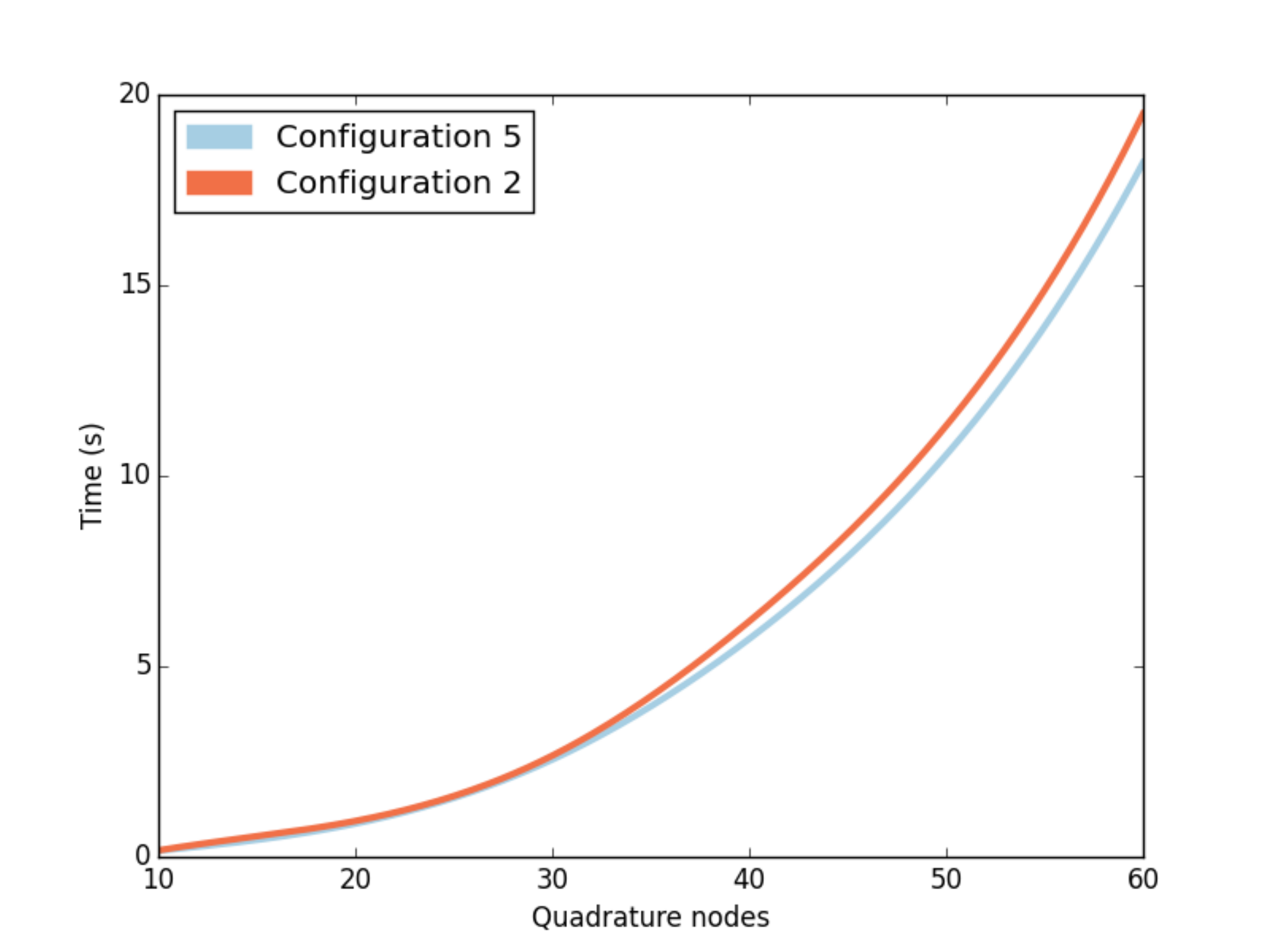}}
\subfigure[Particular]{\includegraphics[width=0.20\textwidth]{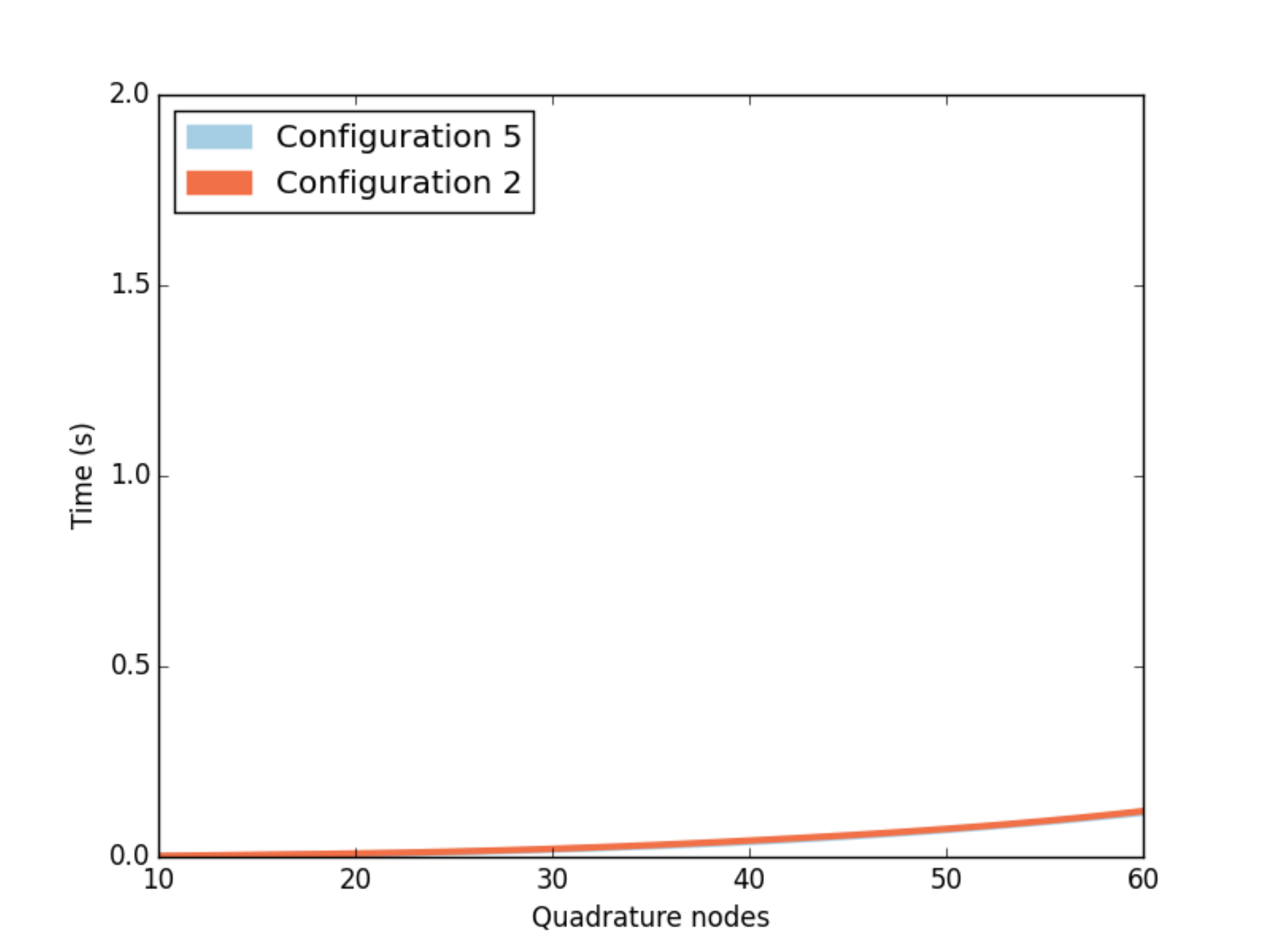}}\\
\subfigure[Boundaries]{\includegraphics[width=0.20\textwidth]{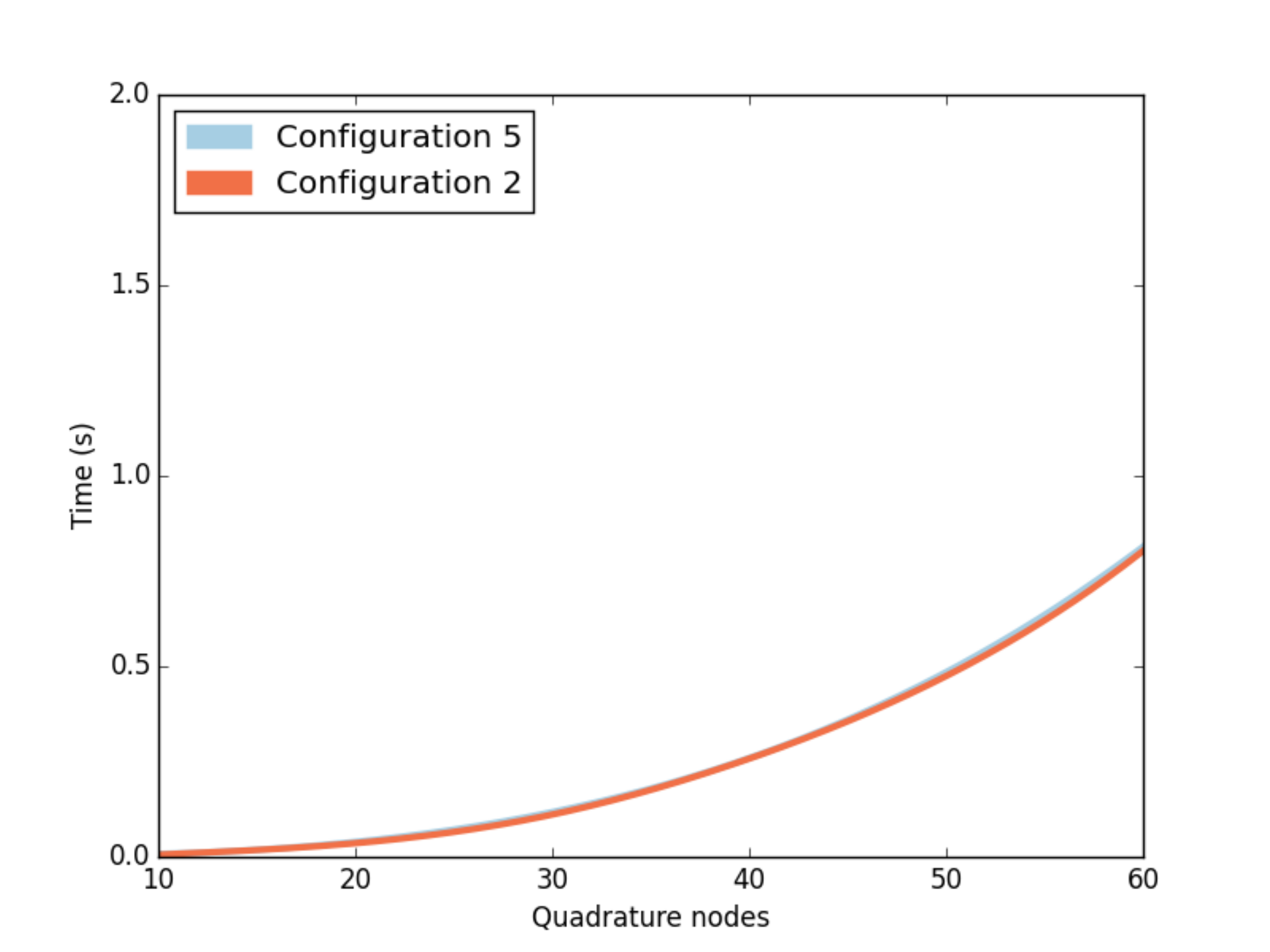}}
\subfigure[Reconstruction]{\includegraphics[width=0.20\textwidth]{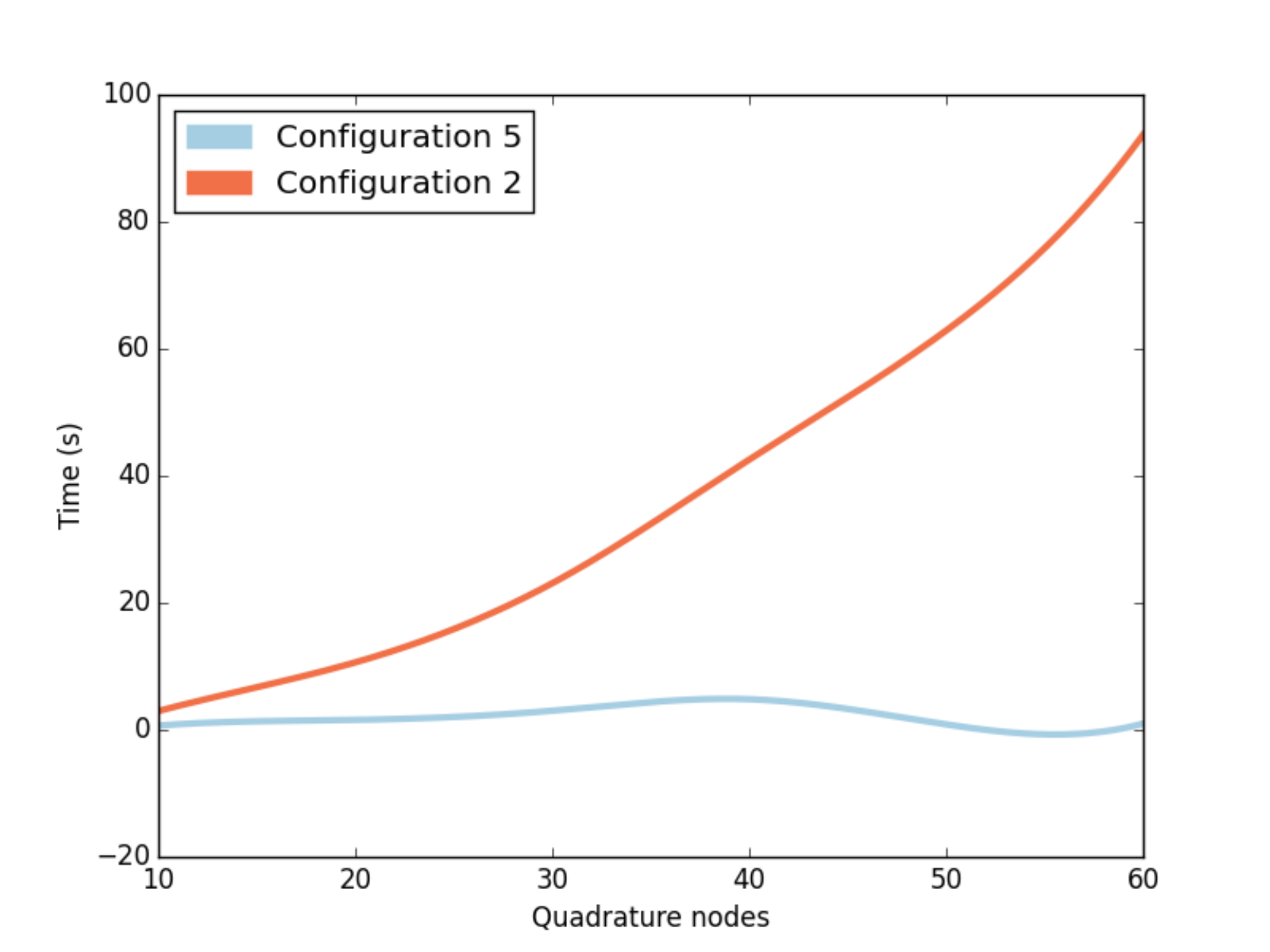}}\\
\caption{Cost of each solution step for different quadrature sizes.}
\label{img_quadrature_2}
\end{figure}

The quadrature size improves the accuracy of the DOM computation. However, they also increase the computation time. In fact, according to the analysis carried out in the previous section, the computation speed should decrease quadratically or cubically (depending on the computation) with the increase in the quadrature size $N$. 
Figure \ref{img_quadrature_1} plots total computation time as a function of quadrature size and figure \ref{img_quadrature_2} plots the breakup of the computation time for execution of each step. The curves are consistently in agreement with our expectation. As discussed earlier, except for the reconstruction step, the computation times for the remaining steps were almost the same for the sequential and the parallel configuration. The reconstruction step of the parallel implementation (configuration 5) did not seem to be affected by the increase in in quadrature size. This could be because the problem size used in the experiments were still not enough to make a full utilization of GPU computation units. Overall, because of the accelerated reconstruction step, the speed of the sequential implementation (configuration 2) lagged behind in parallel implementation (configuration 5) by more than a constant factor.

\begin{figure}[h]
\centering
\includegraphics[width=0.43\textwidth]{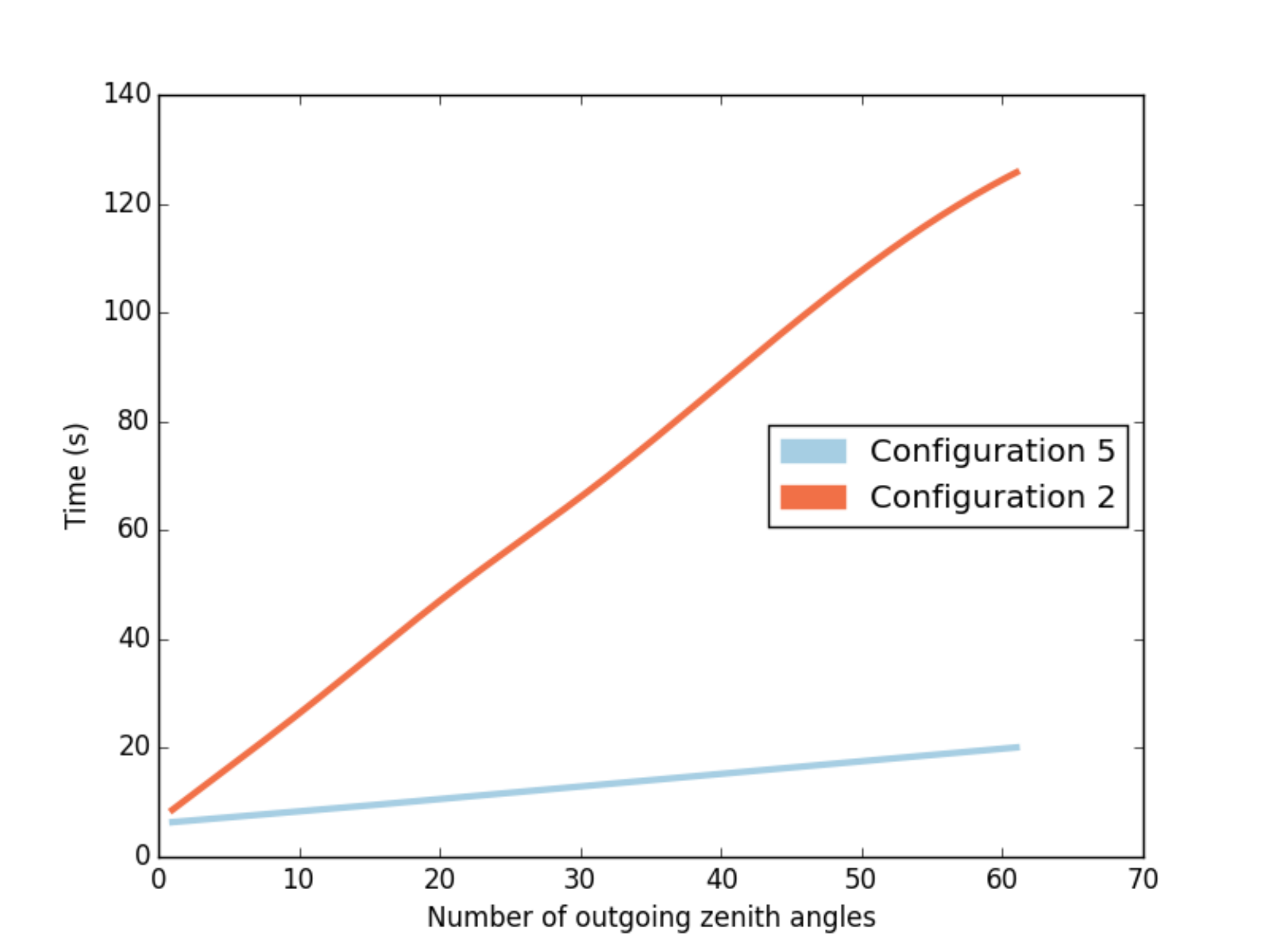}
\caption{Solution cost as a function of the number of outgoing directions.}
\label{img_results_outgoing_1}
\end{figure}

\begin{figure}[h]
\centering
\subfigure[Homogeneous]{\includegraphics[width=0.20\textwidth]{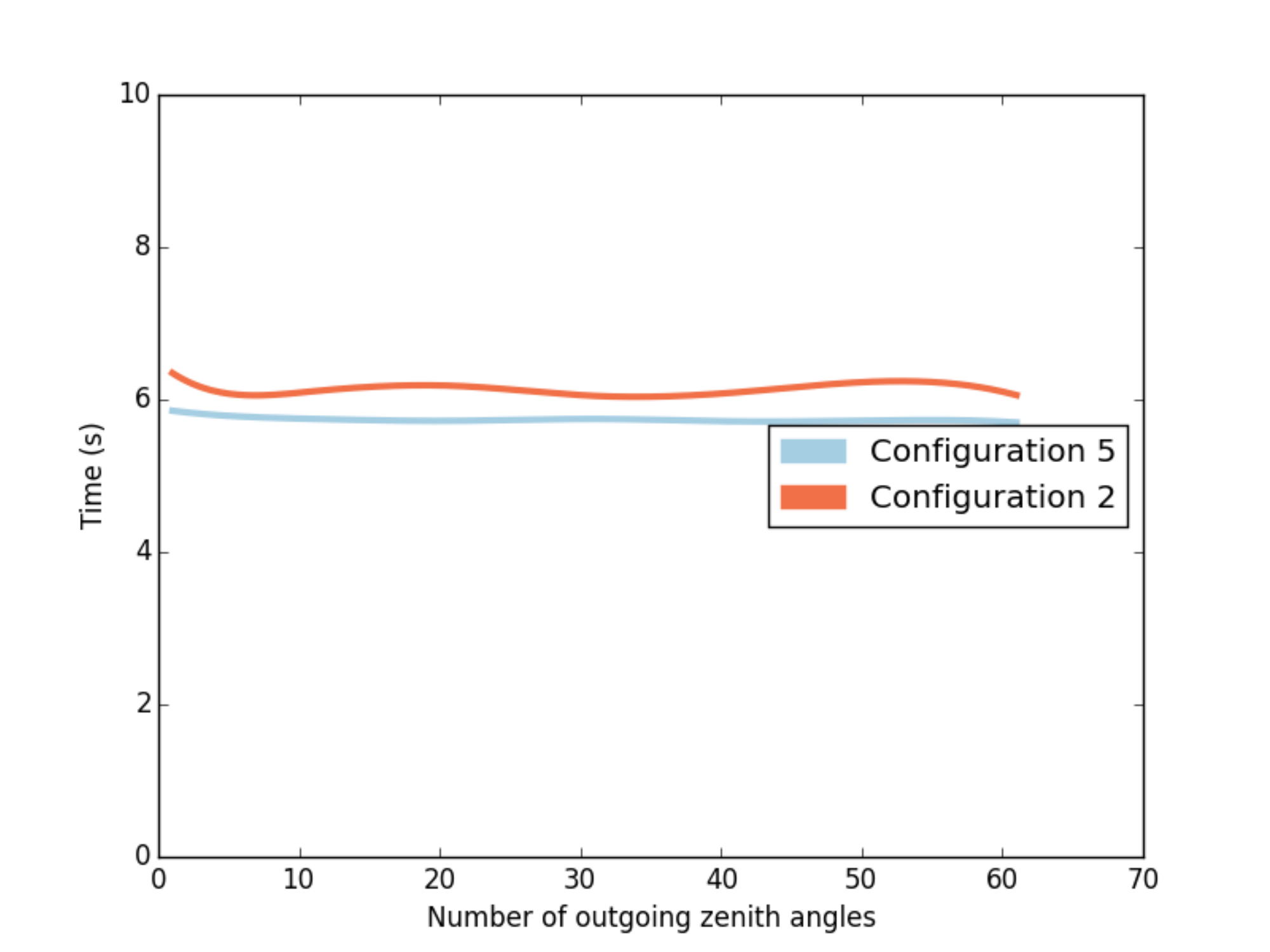}}
\subfigure[Particular]{\includegraphics[width=0.20\textwidth]{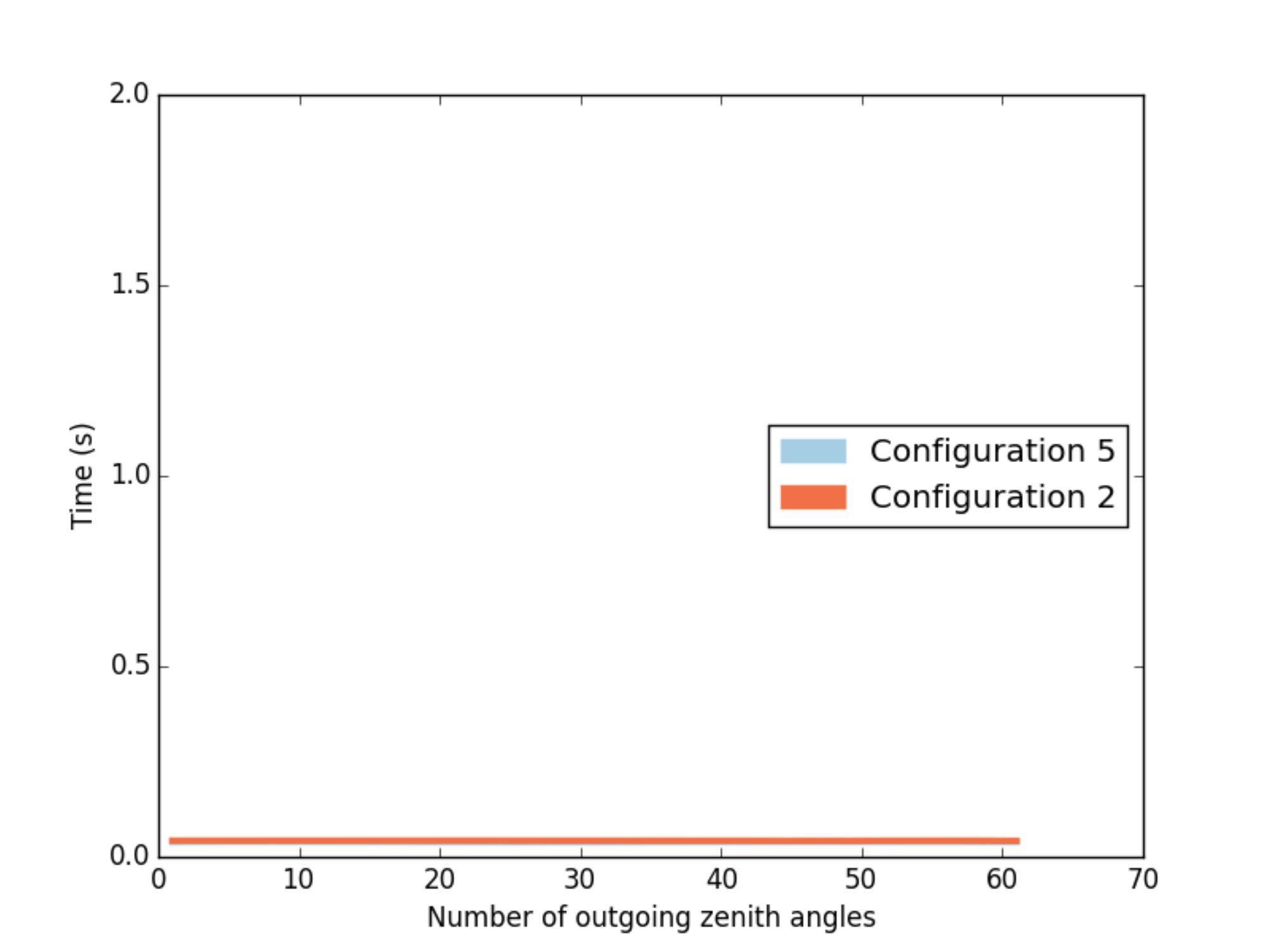}}\\
\subfigure[Boundaries]{\includegraphics[width=0.20\textwidth]{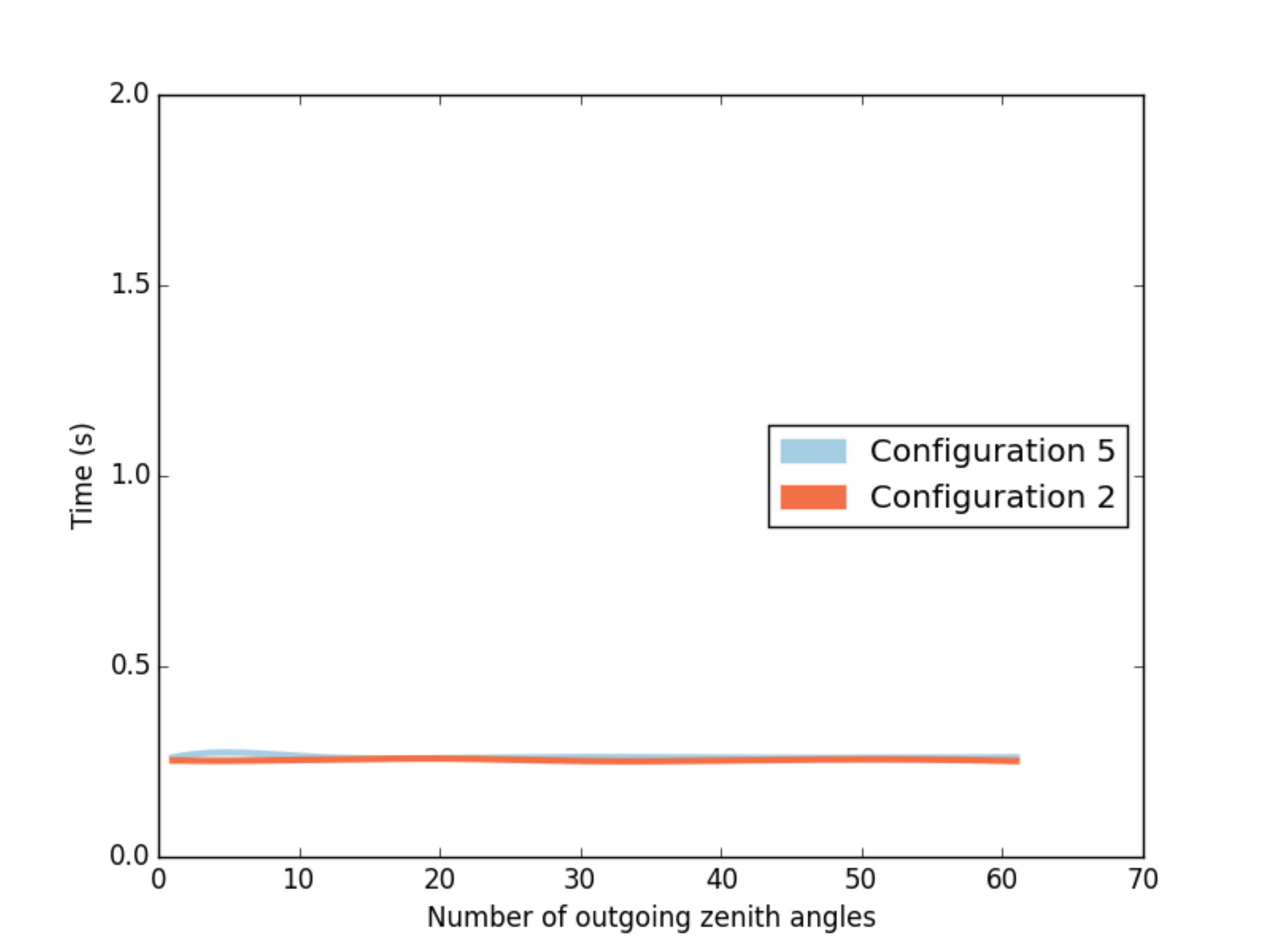}}
\subfigure[Reconstruction]{\includegraphics[width=0.20\textwidth]{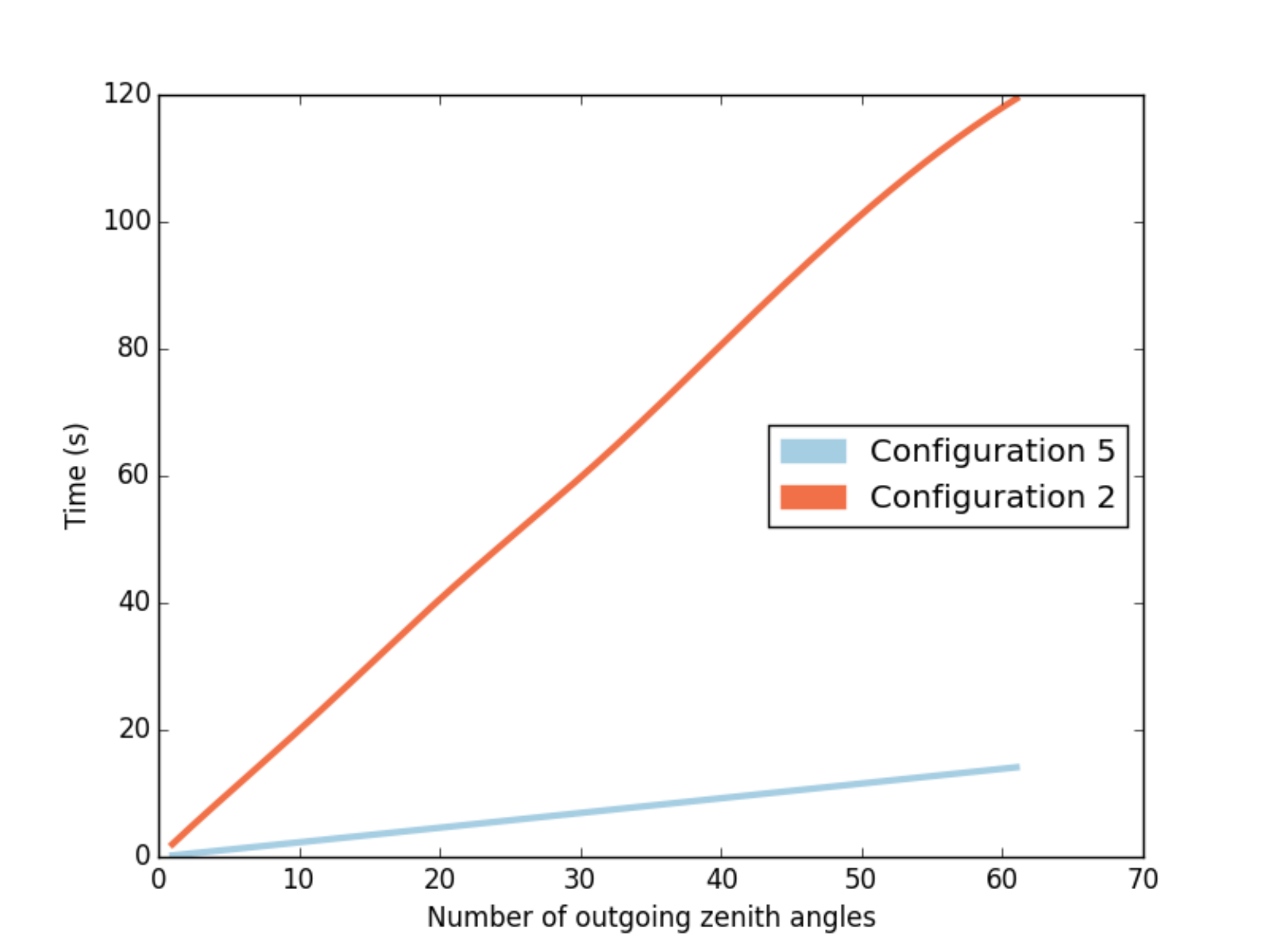}}\\
\caption{Cost of each solution step for different numbers of outgoing directions.}
\label{img_results_outgoing_2}
\end{figure}

\subsubsection {Influence of the number of direction of the radiance field}
The only step affected by the number of outgoing directions ($\mu$) of the radiance field is the reconstruction step, and the cost analysis shows a linear relationship between this number and the reconstruction cost. The plots in figures \ref{img_results_outgoing_1} and \ref{img_results_outgoing_2} agree with our analysis. The computation time of all the steps except the reconstruction step were independent of the number of outgoing directions. As observed earlier the  reconstruction step benefited the most from the reconstruction, and hence did the total computation time.

\begin{figure}[h]
\centering
\includegraphics[width=0.43\textwidth]{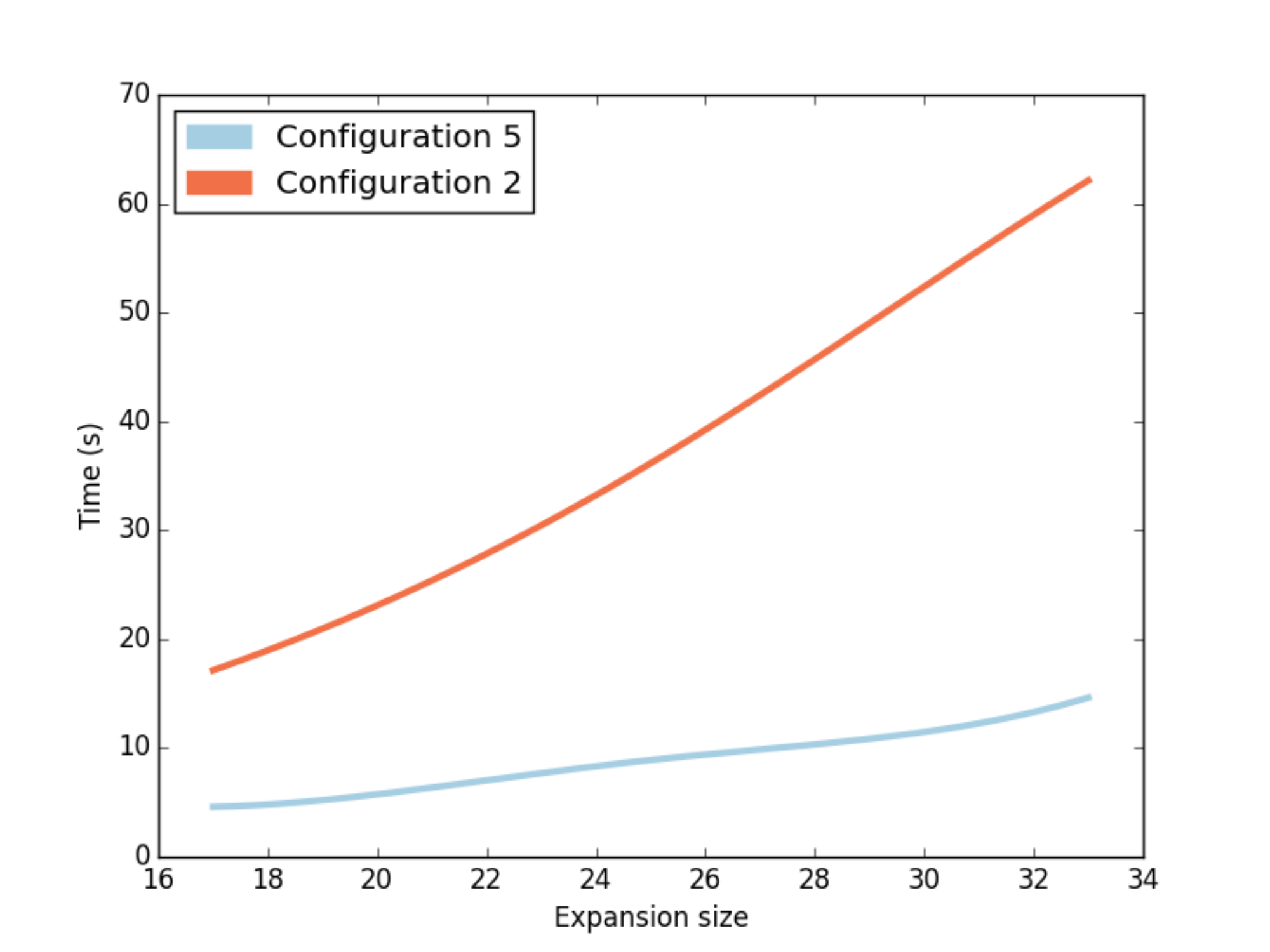}
\caption{Solution cost as a function of the expansion size.}
\label{img_expansion_1}
\end{figure}

\begin{figure}[h]
\centering
\subfigure[Homogeneous]{\includegraphics[width=0.20\textwidth]{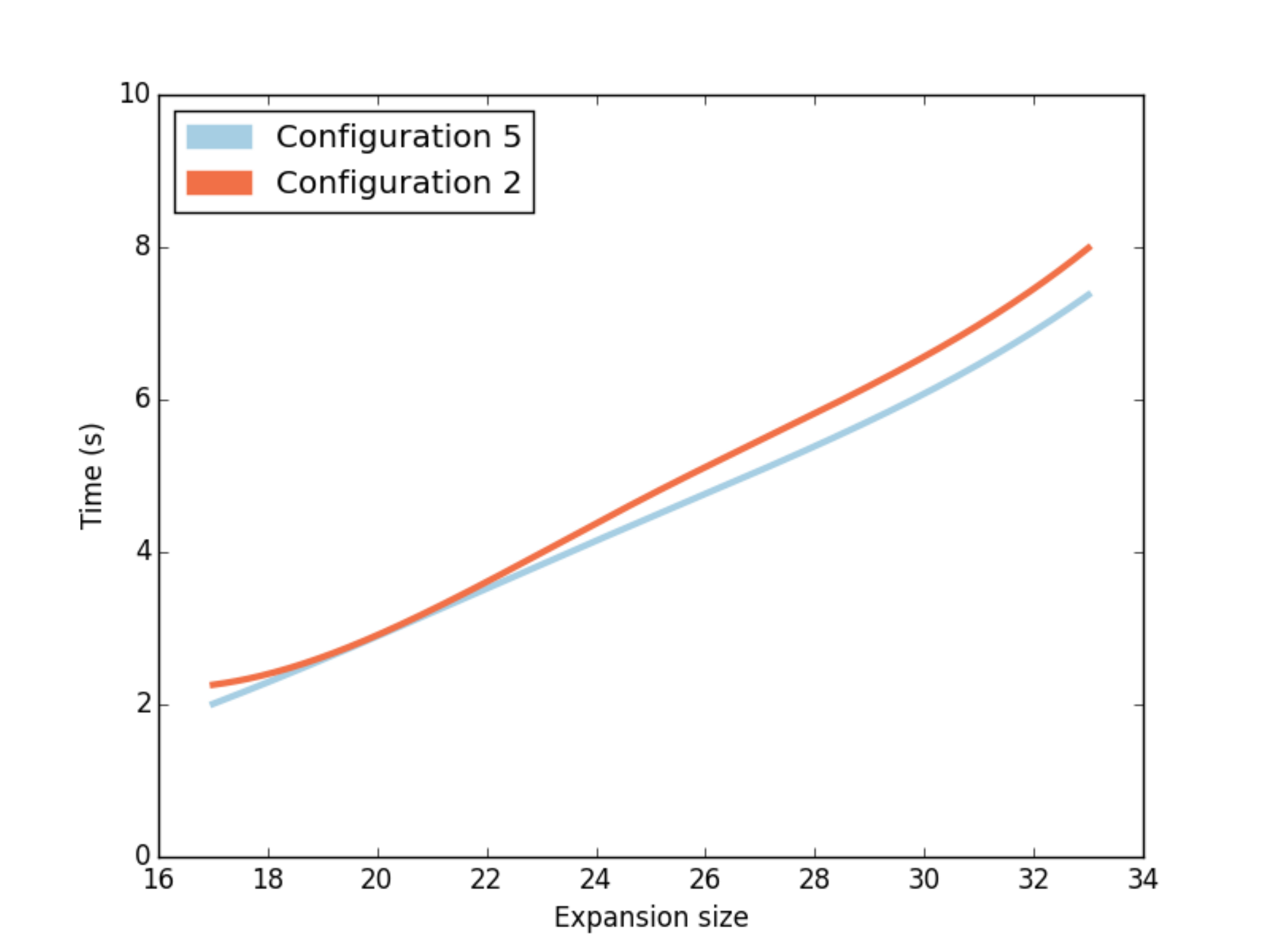}}
\subfigure[Particular]{\includegraphics[width=0.20\textwidth]{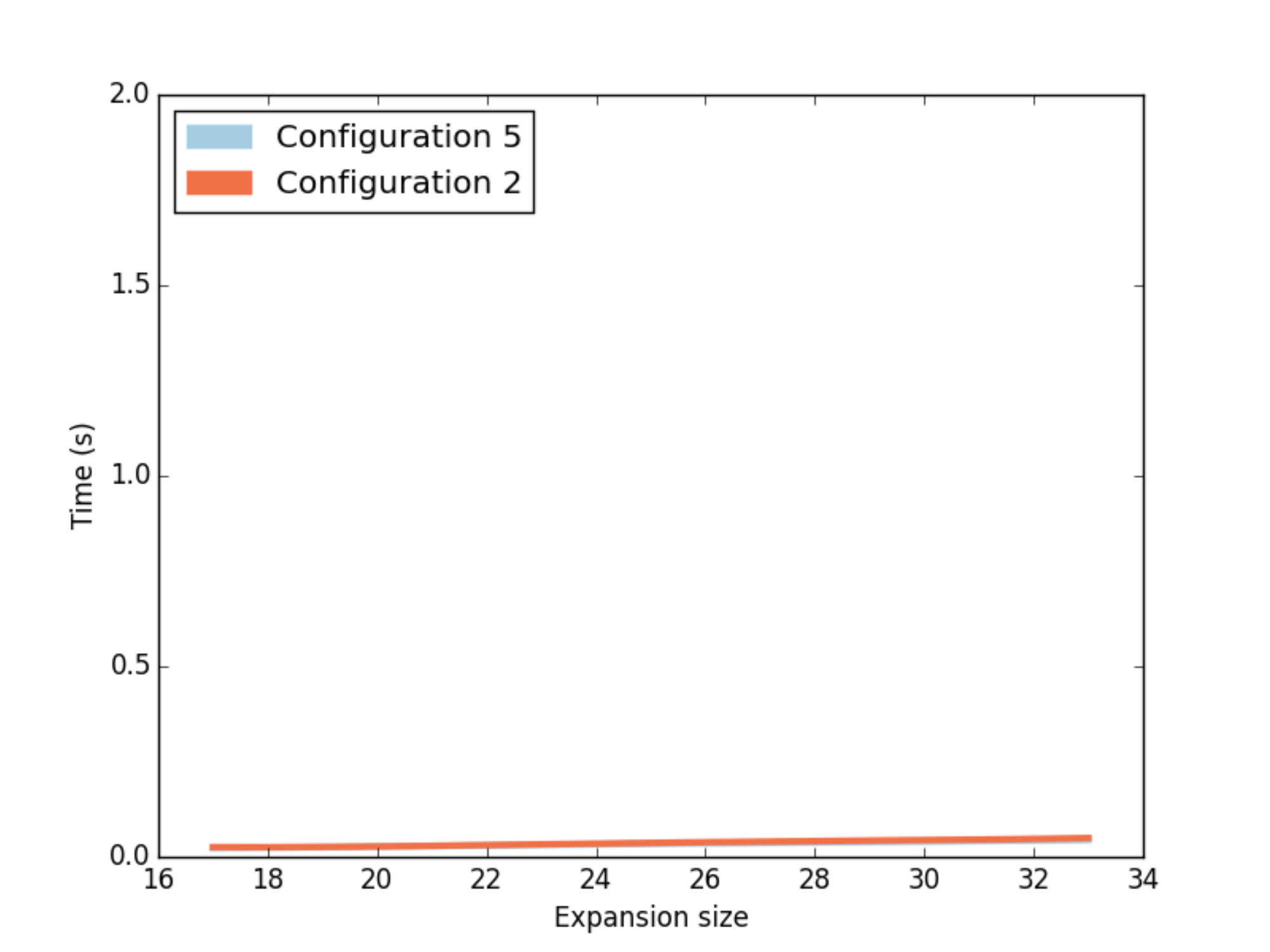}}\\
\subfigure[Boundaries]{\includegraphics[width=0.20\textwidth]{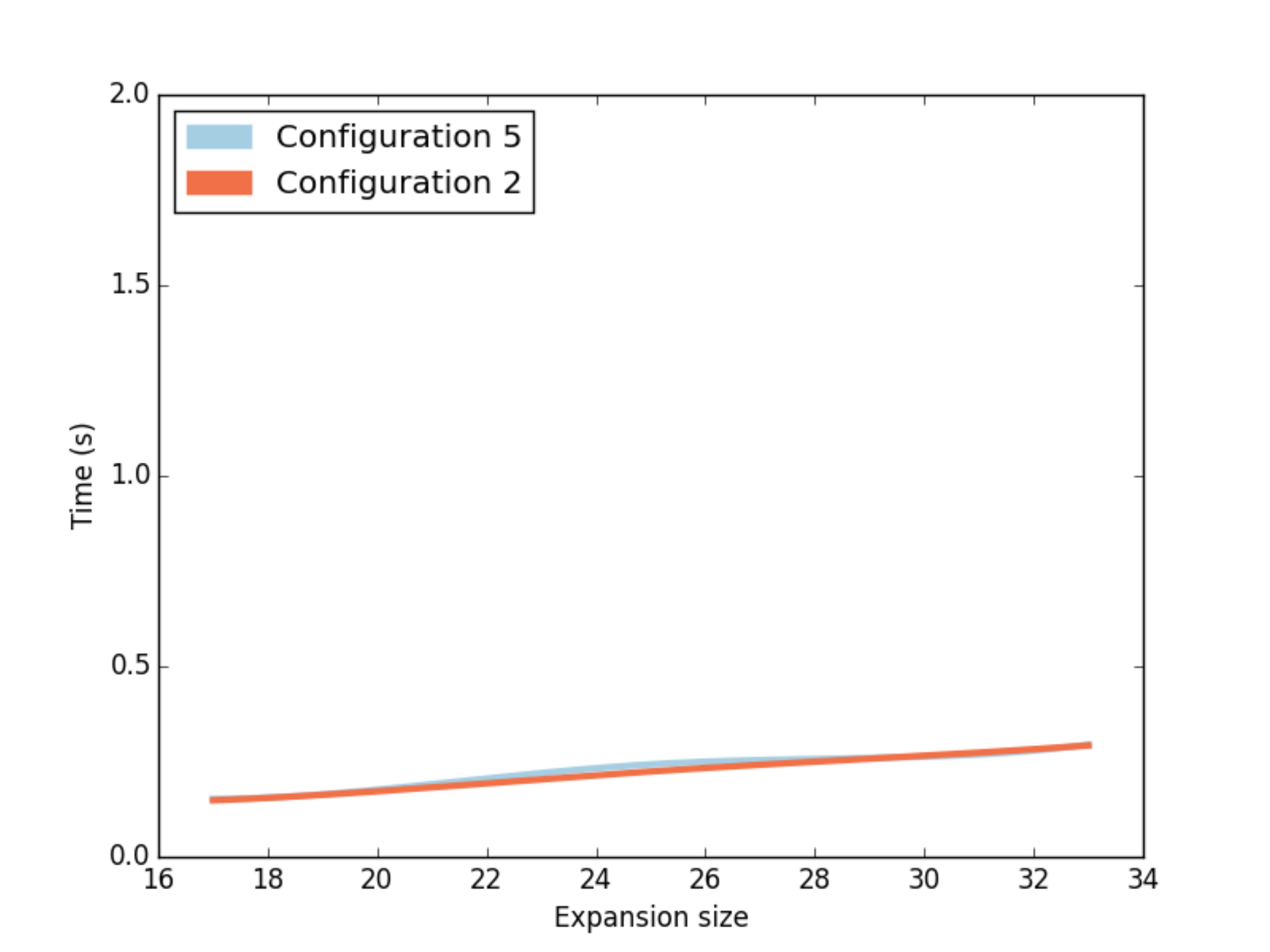}}
\subfigure[Reconstruction]{\includegraphics[width=0.20\textwidth]{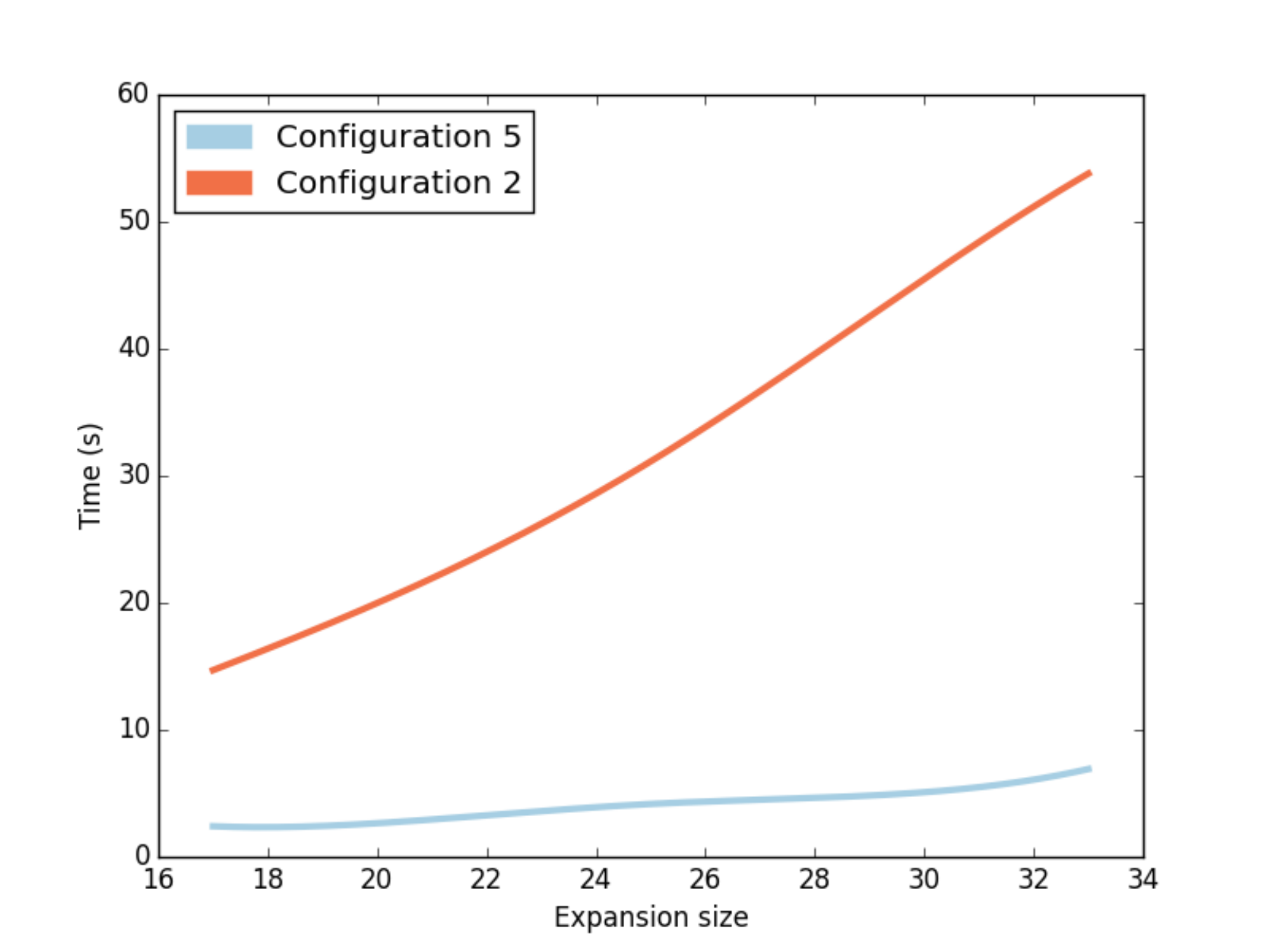}}\\
\caption{Cost of each solution step for different expansion sizes.}
\label{img_expansion_2}
\end{figure}

\subsubsection{Influence of the order of expansion}

The order $L$ of the Fourier expansion of the phase function linearly affects the number of times the computation steps are executed. So the computation time is expected to vary linearly with $L$. 
Figures \ref{img_expansion_1} and \ref{img_expansion_2} plot the total computation time and computation times of the individual steps respectively, as a function of the order of expansion\footnote{The order of the Fourier expansion of the phase function depends on the material. For particles of a given material and a given wavelength of light, it obviously depends on the particle size. So to experiment with different orders of expansion, we varied the size of the particles (between $0.2\mu m$ and $0.7\mu m$) in the layer and then sorted the phase functions by their order of expansion.}.
 The plots show linear relation between the order of expansion and the total computation time. However, as in earlier observations, the parallel configuration did not perform any better for the computation of the homogeneous, particular and boundary steps as compared to the sequential configuration.

\subsection{BRDF computation}
\begin{figure}[h]
\centering
\includegraphics[width=0.43\textwidth]{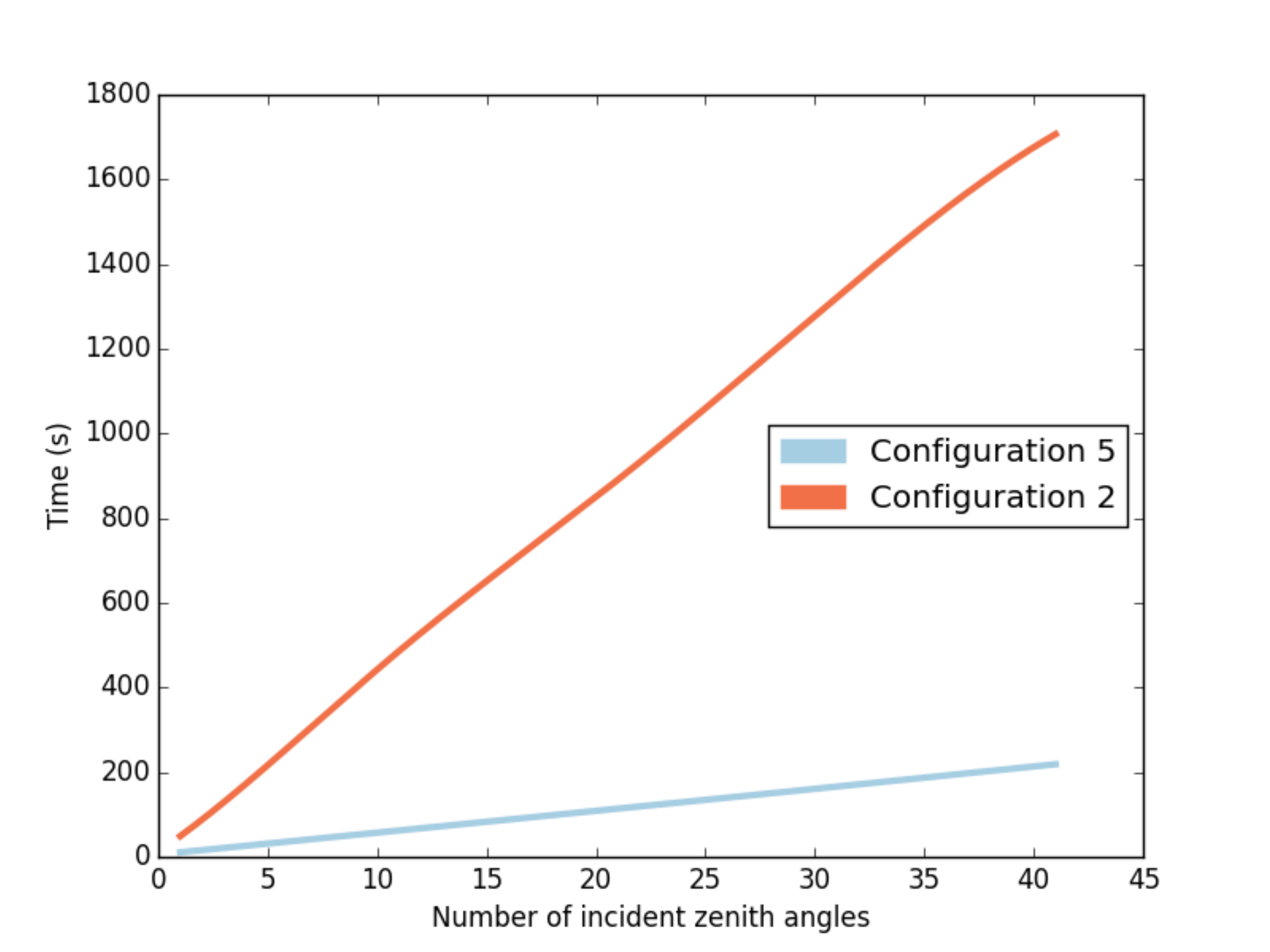}
\caption{Solution cost as a function of the number of incident directions.}
\label{img_results_incident_1}
\end{figure}

\begin{figure}[h]
\centering
\subfigure[Homogeneous]{\includegraphics[width=0.20\textwidth]{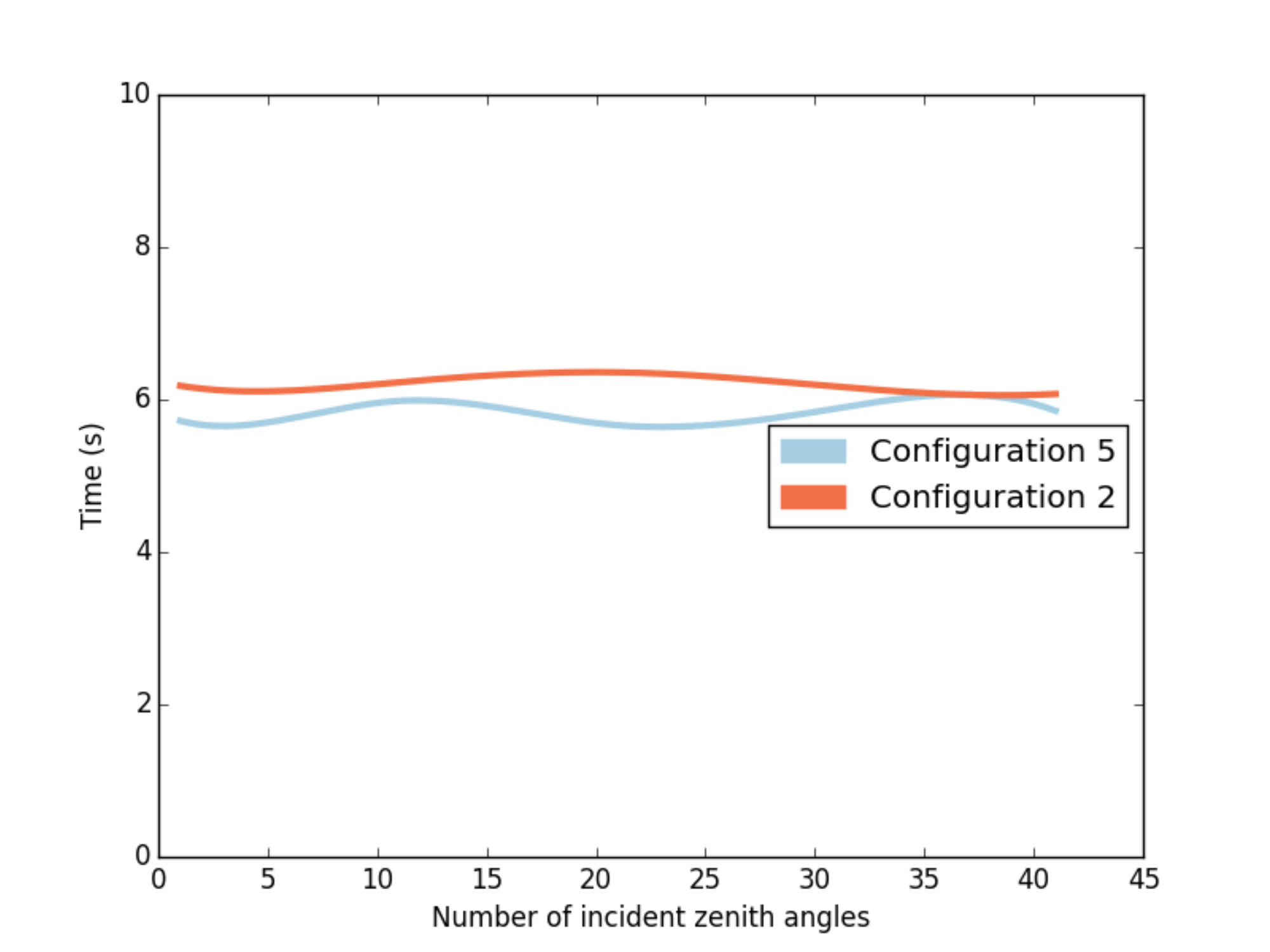}}
\subfigure[Particular]{\includegraphics[width=0.20\textwidth]{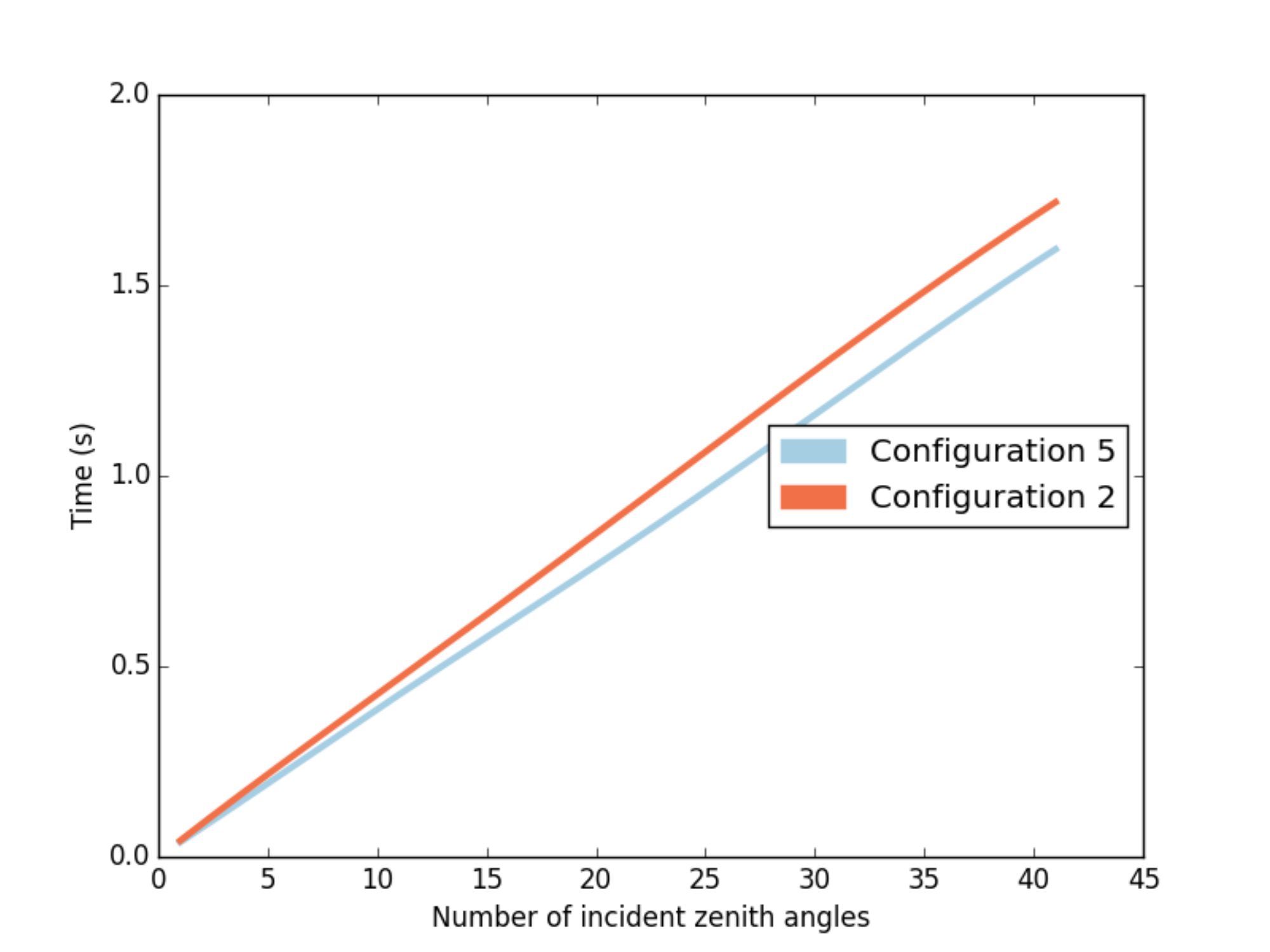}}\\
\subfigure[Boundaries]{\includegraphics[width=0.20\textwidth]{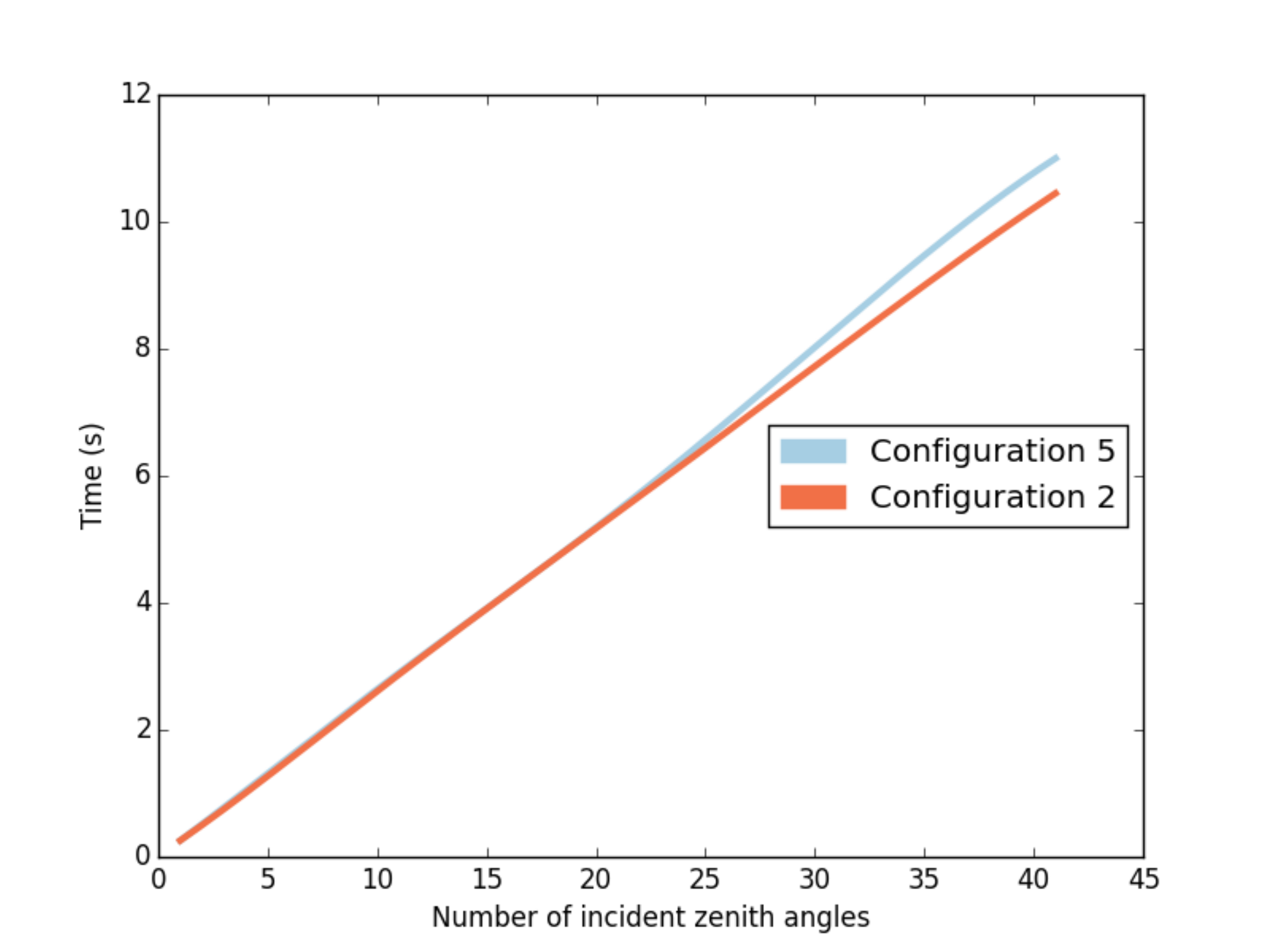}}
\subfigure[Reconstruction]{\includegraphics[width=0.20\textwidth]{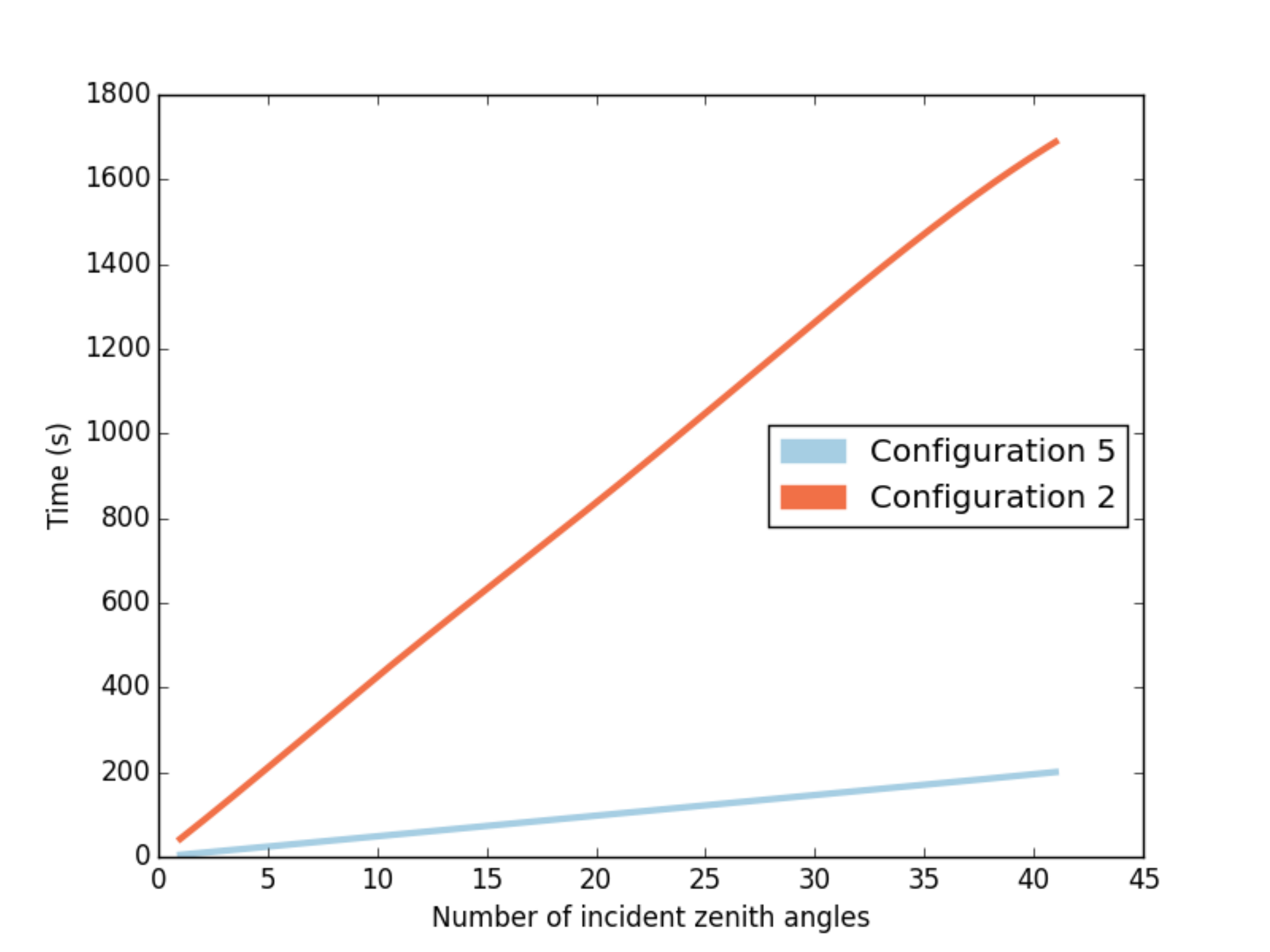}}\\
\caption{Cost of each solution step for different numbers of incident directions.}
\label{img_results_incident_2}
\end{figure}

Finally we show the computation times for BRDF computation. For the BRDF computation VRTE is solved for a number of incident directions. Note that polarized BRDF is a Mueller matrix, so we need a minimum of four incident Stokes vectors for every direction. 
All the parameters affecting the radiance field computation equally affect the BRDF computation. 
We repeated the experiment with the same set up as in section 4.B and measured the computation time as a function of the number of incident directions. Since all the steps except the step involving homogeneous solution had to be repeated for each incident directions and each incident Stokes vector, we expected the cost to increase linearly for those steps with the increase in the number of incident directions.
Figures \ref{img_results_incident_1} and \ref{img_results_incident_2} plot the computation time as a function of the number of incident directions. 
The curves very much agree with the expectation: the computation time of the homogeneous solution step remained unchanged and all the other times showed linear trend.

\subsection{Renderings}
\begin{figure*}[h]
\centering
\subfigure[$Au$ - Particle size: $0.6\mu m$]{\includegraphics[width=0.77\textwidth]{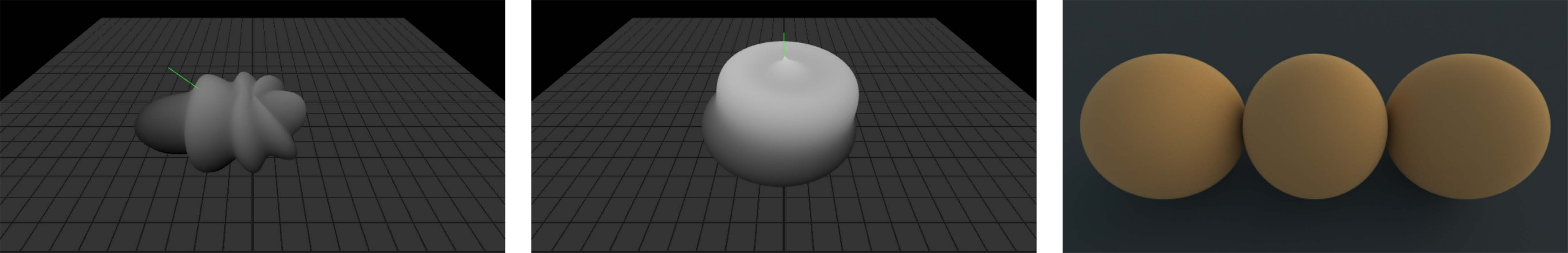}}\\
\subfigure[$Ag$ - Particle size: $0.3\mu m$]{\includegraphics[width=0.77\textwidth]{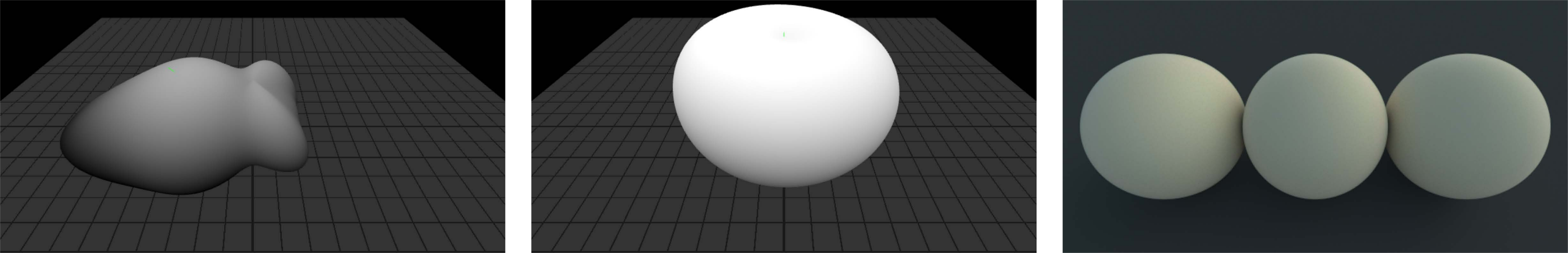}}\\
\subfigure[$AlGaAs$ - Particle size: $0.3\mu m$ (Lobe scale factor = $0.5$)]{\includegraphics[width=0.77\textwidth]{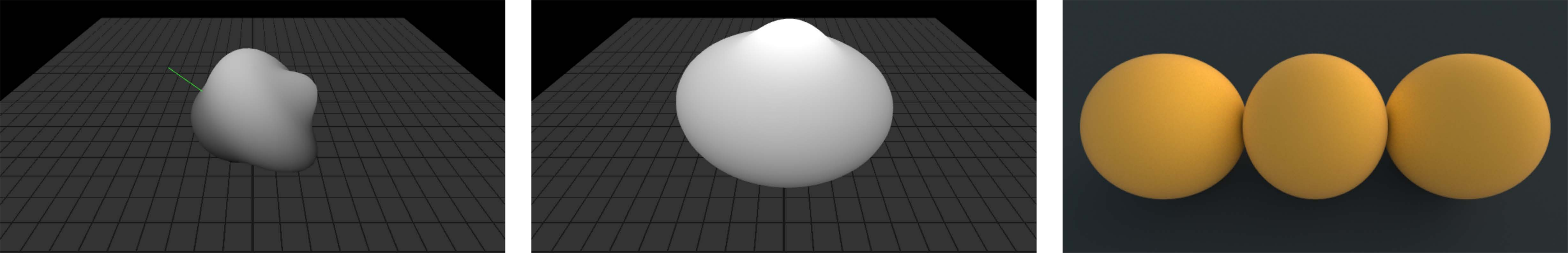}}\\
\subfigure[$AlCu$ - Particle size: $0.3\mu m$]{\includegraphics[width=0.77\textwidth]{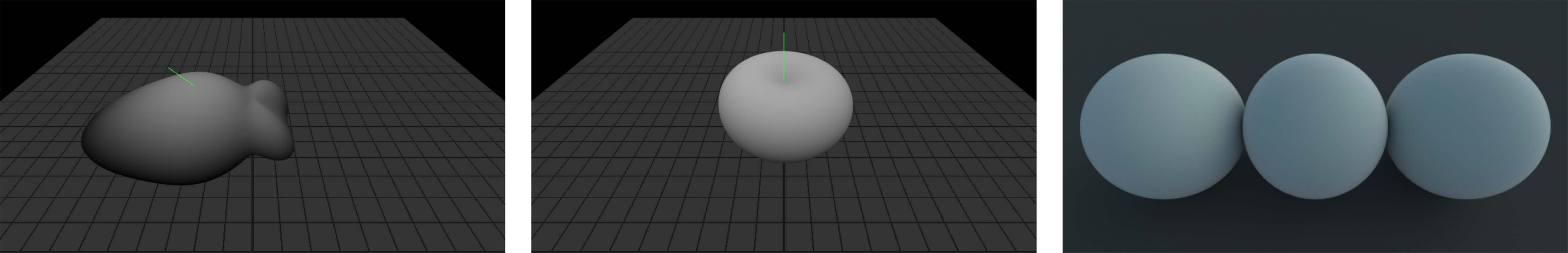}}\\
\subfigure[$ZnSCuB$ - Particle size: $0.8\mu m$ (Lobe scale factor = $0.5$)]{\includegraphics[width=0.77\textwidth]{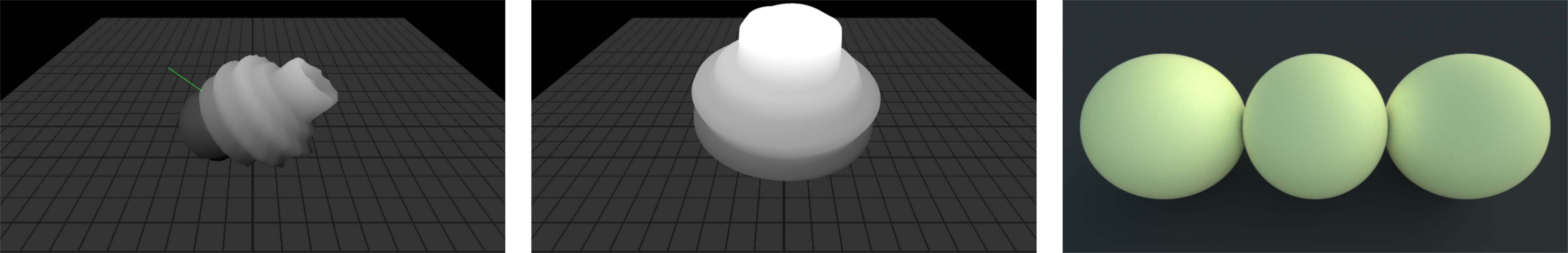}}\\
\subfigure[$PbS$ - Particle size: $0.3\mu m$ (Lobe scale factor = $2.0$)]{\includegraphics[width=0.77\textwidth]{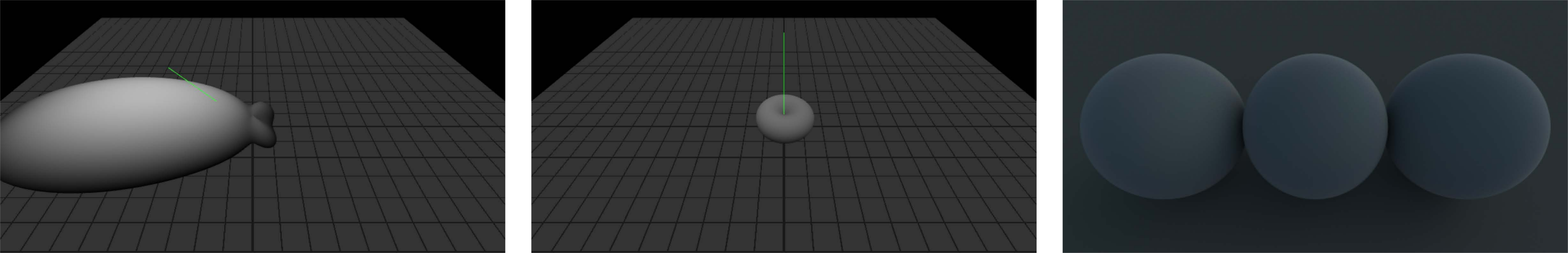}}\\
\subfigure[$Cu$ - Particle size: $0.6\mu m$]{\includegraphics[width=0.77\textwidth]{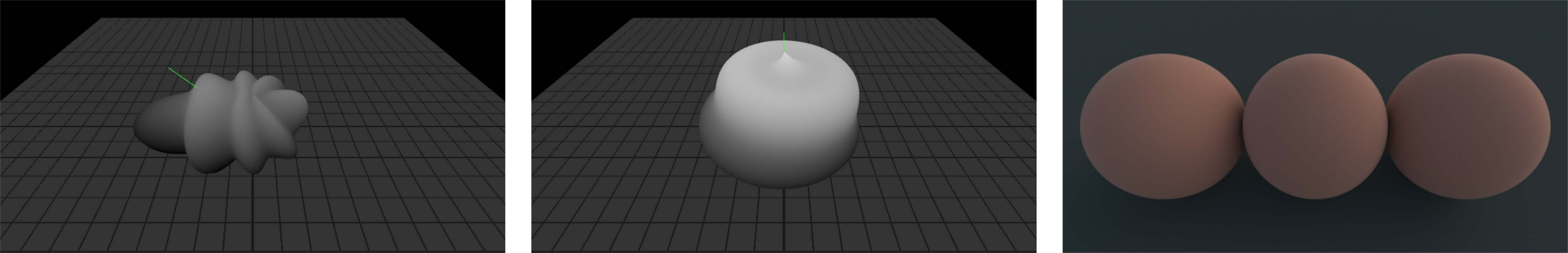}}\\
\subfigure[$TiO_2$ - Particle size: $0.6\mu m$ (Lobe scale factor = $0.5$)]{\includegraphics[width=0.77\textwidth]{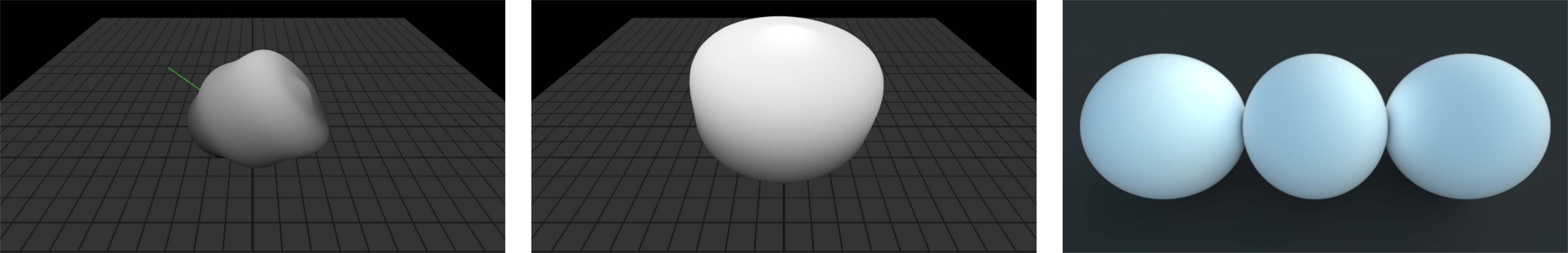}}\\
\caption{BRDF lobes and renderings for various materials and particle size. Lobes are shown for a wavelength $\lambda=640nm$. Left column corresponds to an incident direction $\mu_i = 0.6$, middle column to a normal incident direction ($\mu_i = 1.0$). Right column shows a rendering using 9 wavelengths. Some of the lobes have been scaled to a reasonable size. For those lobes, the scale factor is given below the lobes.}
\label{img_results_brdf_1}
\end{figure*}

We used our solver to compute the BRDF for material layers composed of several particle types. 
Figure \ref{img_results_brdf_1} shows for each material two computed BRDF lobes: one for an oblique incidence ($\mu_0=0.6$) and the other for normal incidence, and a rendered image of a scene composed of spherical objects with the computed BRDF as their surface properties. The scene was illuminated with synthetic skylight. Each row of the figure corresponds to a different material.

The current version of our parallel solver (configuration 5) computes a polarized subsurface BRDF in \~15 minutes, where as the sequential solver (configuration 2) takes close to two hours for this computation.

We have made the solver used in this section available \cite{pivert} for public use. The solver takes the material specification as input, and based on the user's choice computes the polarized radiance field at any layer thickness ($\tau$ value) or computes the polarized BRDF. The output is made available to the user in a tabular form for download. The page also provides a renderer to visualize the computed radiance field or the BRDF lobe. In the latter case, the renderer allows interactive viewing of the computed lobe as a function of incident direction. The rendered sphere images used in this section were obtained using a polarized path tracer. We have also made this path tracer available for public use \cite{pirate}.

\section{Conclusion}
\label{section_conclusion}
In this paper we proposed parallelization of DOM based VRTE solvers for computing polarized light transport inside homogenous layered materials for computing polarized radiance field and for computing polarized subsurface BRDF. We analyzed the cost of each step of the solution, identified potential parallelization steps, and implemented those steps in OpenCL to run them in parallel in GPU. The major bottleneck was found to be in solving eigen problem and in computing matrix inversion. Though we attempted to use GPU based library for these computations, for our problem size we found only minor speedup for matrix inversion computation and minor setback for eigen solution computation. The parallelization gave us significant speedup in the final reconstruction step that dominated the cost of radiance field computation and more so in BRDF computation using the solver. So the overall computation speed-up for our parallel solver was found to be significant as compared to its sequential counterpart.  Our parallel software configuration (that uses a sequential linear algebra library for eigen problem and linear system solution) is about seven times faster than the sequential configuration for BRDF computation with a reasonable number of incident and outgoing directions. 

Though our parallel solver allows us to compute BRDF faster, we believe that we will be able to further improve its speed. 
We are still using an external Linear Algebra library and computing the expensive matrix problems (eigen solution and matrix inversion) sequentially, and furthermore solving them sequentially for each order of expansion. 
In future we plan to write our parallelized eigen solver and linear system solver that will allow us to compute eigen solution and matrix inversion for all the expansion orders in one function call each. This will result in an increase in problem size for better utilization of GPU resources, and will help us bring down the computation time further.

\end{document}